\documentclass[a4paper,11pt]{article}
\usepackage{amssymb}

\usepackage{epsfig}
\usepackage{graphicx}
\usepackage{amsmath}

\begin{document}

\pagestyle{empty}

\title{{\bf A study of top polarization in single-top production at the LHC}}
\author{D. Espriu$^{a,b}$\thanks{%
espriu@ecm.ub.es} \ and J. Manzano$^b$\thanks{%
manzano@ecm.ub.es} \\
\\
$^a$CERN, TH Division, 1211 Geneva 23\\
\\
$^b$Departament d'Estructura i Constituents
de la Mat\`{e}ria\\
and\\
CER for Astrophysics, Particle Physics and Cosmology,\\
Universitat de Barcelona, \\
Diagonal, 647, E-08028 Barcelona}
\date{}
\maketitle

\begin{abstract}
This paper complements the study of single top production at the LHC
aiming to estimate the sensitivity of
different observables to the magnitude of the effective couplings.
In a previous
paper the dominant $W$-gluon fusion mechanism
was considered, while here we extend the analysis to the subdominant
(10\% with our set of experimental cuts) $s$-channel process.
In order to distinguish
left from right effective couplings it is required to
consider polarized cross-sections and/or include $m_b$
effects. The spin of the top
is accessible only indirectly by measuring the angular distribution
of its decay products. We show that the presence of effective
right-handed couplings implies necessarily that the top is not
in a pure spin state. We discuss to what extent quantum interference
terms can be neglected in the measurement and therefore simply multiply
production and decay probabilities clasically. The coarsening involved in the
measurement process makes this possible. We determine for each process
the optimal spin basis where theoretical errors are
minimized and, finally, discuss
the sensitivity in the $s$-channel to the effective right-handed coupling.
The results presented here are all analytical and include $m_b$ corrections.
They are derived within the narrow width approximation for the top.

\end{abstract}

\vfill
\vbox{
UB-ECM-PF 02/18\null\par
September 2002\null\par
}

\clearpage

\section{Introduction}
\label{intro}

At present not a lot is known about the
$Wt\bar{b}$ effective coupling. This is perhaps
best evidenced by the fact that the
current experimental results for the (left-handed) $K_{tb}$ matrix element
give \cite{PDG}
\begin{equation}
\frac{|K_{tb}|^{2}}{|K_{td}|^{2}+|K_{ts}|^{2}+|K_{tb}|^{2}}=0.99\pm 0.29.
\end{equation}
In the Standard Model this matrix element is expected to be close to 1. It
should be emphasized that these are the `measured' or `effective' values of
the CKM matrix elements, and that they do not necessarily correspond, even
in the Standard Model, to the entries of a unitary matrix on account of the
presence of radiative corrections. These deviations with respect to unitary
are expected to be small ---at the few per cent level at most--- unless new
physics is present and makes an unexpectedly large contribution.
At the Tevatron the left-handed couplings are expected
to be eventually measured with a 5\% accuracy \cite{TEVA}.

As far as experimental bounds for the right handed effective
couplings is concerned, the more stringent ones come at present
from the measurements on the $b\rightarrow s\gamma $ decay at CLEO
\cite{cleo}. Due to a $m_{t}/m_{b}$ enhancement of the chirality
flipping contribution, a particular combination of mixing angles
and effective right-handed couplings  can be bound very precisely.
The authors of \cite {LPY} reach the conclusion that
$|\mathrm{Re}(g_R)|\leq 0.4\times 10^{-2}$. However, considering
$g_R$ as a matrix in generation space, this bound only constraints
the $tb$ element. Other effective couplings involving the top
remain virtually unrestricted from the data. The previous bound on
the right-handed coupling is a very stringent one. It should be
obvious that the LHC will not be able to compete with such a
bound. Yet, the measurement will be a direct one, thus ruling out
some contrived models where substantial cancellations might
hypothetically avoid the $b\to s\gamma$ constraint. For the value
of the effective couplings in some specific models see e.g.
\cite{belyaev}.

\begin{figure}[!hbp]
\begin{center}
\includegraphics[width=8cm]{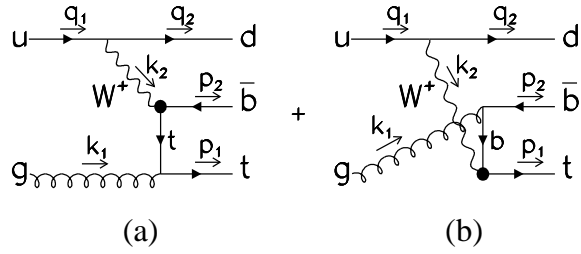}
\end{center}
\caption{Feynman diagrams contributing single top production subprocess. In
this case we have a $d$ as spectator quark}
\label{u+gt+b-d+tot}
\end{figure}

\begin{figure}[!hbp]
\begin{center}
\includegraphics[width=7.5cm]{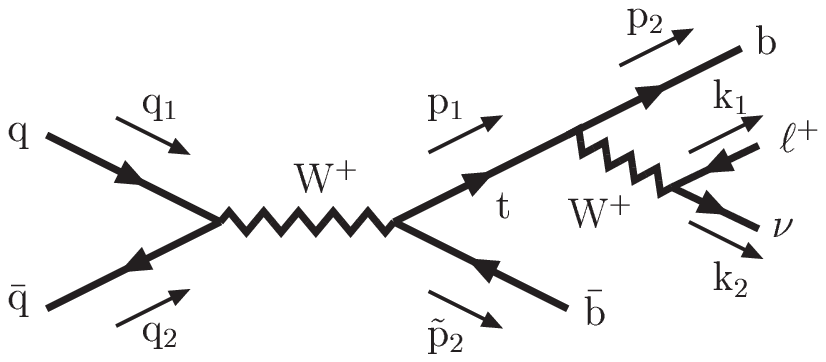}
\end{center}
\caption{Feynman diagram contributing to single top production in
the subdominant $s$-channel process. The top decay is also shown
in this figure} \label{singletopschannelanddecay}
\end{figure}

At LHC energies the mechanism underlying single top production, therefore
allowing a direct test of the $W\bar{b}t$ effective couplings $g_L$ and
$g_R$,  consists
of several different processes (see e.g. \cite{Tait}).
The dominant process is the
so-called $W-$gluon fusion channel, or $t$-channel process. The electroweak
subprocesses corresponding to this channel are depicted in Fig. \ref
{u+gt+b-d+tot}, where light $u$-type quarks or $%
\bar{d}$-type antiquarks are extracted from the protons.
Besides this dominant channel (250 pb at LHC \cite{SSW}) single tops are
also produced through the process where the $W^{+}$ boson interacts with a $%
b $-quark extracted from the sea of the proton (50 pb)\cite{SSW}
and in the quark-quark fusion or $s$-channel process (10 pb) which
is depicted in Fig.\ref{singletopschannelanddecay}.
The numbers quoted here correspond to total cross-sections. The
separation between the sub-dominant processes and the dominant
$W$-gluon fusion is purely kinematical\cite{SSW,EM}. By placing a
cut on the $p_T$ of the detected $\bar{b}$ quark, the former
process can be eliminated altogether. This also eliminates a
sizeable fraction of the tops produced via the $W$-gluon fusion
mechanism (about two thirds for the cuts we use). The cut on $p_T$
has the additional bonus of making the QCD corrections manageable.
One is therefore left with those single tops coming from the
$W$-gluon fusion mechanism ($t$-channel) and the subdominant
${s}$-channel process. The later one is actually the main object
of our interest in this article, although we will also have many
comments to make on the $t$-channel process.

In a proton-proton collision a bottom-anti-top pair is also produced through
analogous subprocesses. The analysis of such anti-top production processes
is similar to the top ones and the corresponding cross sections can be
easily derived doing the appropriate changes.

In a previous paper\cite{EM} we have analyzed the sensitivity of
different LHC observables to the magnitude of the charged current
effective couplings considering only the dominant $W$-gluon fusion
channel. In that work we did not consider the subsequent decay of
the top in any detail. We did, however, a complete analytical
calculation of the subprocess cross sections, for general left and
right effective couplings and including bottom mass corrections. A
$p_{T}>30$ GeV cut in the transverse momentum of the produced
$\bar{b}$ quark was implemented in \cite{EM} and, accordingly,
only the so-called $2\rightarrow 3$ process was retained,
excluding top production off a $b$-quark from the proton Fermi
sea. Given the (presumed) smallness of the right handed couplings,
the bottom mass plays a role which is more important than
anticipated, as the mixed
crossed $g_{L}g_{R}$ term, which actually is the most sensitive one to $%
g_{R} $, is accompanied by a $b$ quark mass. The reader is encouraged
to see \cite{EM}, where a very detailed analysis is presented.

Typically the top quark decays weakly well before strong interactions become
relevant, so we could in principle `measure' its polarization
state with virtually
no contamination of strong interactions (see e.g. \cite{parke,mandp} for
discussions this point) and try to establish interesting observables
based on this measurement.
In fact it is not difficult to convince oneself that in order to
disentangle left from right effective couplings, it is
almost compulsory to be able to
`measure' the polarization of the top. This will become apparent
from the formulae presented in section \ref{differentialc}.
For this reason we have
derived  in this work and in\cite{EM} analytical
expressions for
the cross sections for the
production of polarized tops or anti-tops. To this end one
introduces the spin projector
\begin{equation*}
\left( \frac{1+\gamma _{5}\not{n}}{2}\right) ,
\end{equation*}
with
\begin{equation}
n^{\mu } =\frac{1}{\sqrt{\left( p_{1}^{0}\right) ^{2}-\left( \vec{p}%
_{1}\cdot \hat{n}\right) ^{2}}}\left( \vec{p}_{1}\cdot \hat{n},p_{1}^{0}\hat{%
n}\right) ,\qquad \hat{n}^{2} =1,\qquad n^{2}=-1,
 \label{spinfour}
\end{equation}
as the polarization projector for a particle or anti-particle of momentum $%
p_{1}$ with spin in the $\hat{n}$ direction. The calculation of the
subprocess cross sections have been performed in this work and in \cite{EM}
for an arbitrary polarization vector $\hat{n}$.

Obviously, however, the top decays very shortly after
production, so the only practical way one can measure the spin of the top is
through its influence on
the angular distribution of the leptons produced in the decay.
It is tacitly assumed in most of the works published on this subject
that the
decaying top is in a pure spin state for all practical purposes; i.e. its
polarization vector is pointing in a particular direction in space in a
given reference frame.

In the tree-level Standard Model this is not quite true, but it is
almost true. The tree level Standard Model corresponds in our
notation to taking $g_L=1$ and $g_R=0$.  Imposing the a cut on
$p_T$ we have mentioned, only two subprocesses contribute;
$W$-gluon fusion and the $s$-channel process. The later provides
100\% polarized tops in a certain direction (to be discussed
latter). The situation in the $t$-channel process is a bit more
complicated. The results from our previous analysis presented in
\cite{EM} show that single top production is highly, but not
fully, polarized in this case too (84 \% in the optimal basis,
with the present set of cuts). This is a high degree of
polarization, but still well below the 90+ claimed by Mahlon and
Parke in \cite{mandp}. We understand this being due to the
presence of a 30 GeV cut in $p_T$. In fact, if we remove this cut
completely we get
 91 \% polarization, in rough agreement with \cite{mandp}
(note that we do not include the $2\rightarrow 2$  or $b$-sea process).
Inasmuch as they can be compared our results for the tree-level
Standard Model are in good agreement with
those presented in \cite{SSW} in what concerns the total cross-section. These
considerations are quite independent of the choice of the strong
subtraction scale, which is by far the largest
source of uncertainty\footnote{
Since we perform a leading
order calculation in QCD, the scale dependence is large. We have made two
different choices: (a)  $\mu=p_{T}^{cut}$ is used
as scale in $\alpha _{s}$ and the gluon PDF, while the virtuality
of the $W$ boson is
used as scale for the PDF of the light quarks in the proton. This
gives an excellent agreement
with the calculations in \cite{SSW}. (b) $\mu ^{2}=\hat{s}$, $\hat{s}$
being the
center-of-mass energy squared of the $qg$ subprocess. The total cross
section above the cut is then roughly speaking two thirds of the previous
one, but no substantial change in the distributions takes place. This
is the typical error for LO calculations in the present kinematical
regime. The total cross section has been known to NLO for some time \cite{TZS},
while NLO results for the differential cross section have become available
just recently \cite{NLO}}. Let us assume now for the sake of
discussion that the polarization
is indeed 100\% .
The top subsequently decays (say emitting a positively
charged lepton). One can compute the angular probability
distribution of the lepton with respect to the polarization direction
in the Standard Model,
multiply the two probabilities and compare the experimental result
with the theoretical prediction.

In fact things are a lot more subtle. First of all, we have seen
that even in the Standard Model polarization is never 100\% .
Furthermore, it turns out that when $g_{R}\neq 0$, i.e. beyond the
Standard Model,  the top can never be 100\% polarized (see the
discussion in section \ref{differentialc} and in \cite{EM}), not
even in principle.  In other words, the top
is necessarily in a quantum mixed state and is described by a
density matrix. The entries of this density matrix depend on the
momenta of the incoming and outgoing particles; that is to say,
there is an entanglement between spin and momenta.

Of course this complication amounts to a
 small effect because $g_R$ is surely
quite small, even in most models beyond the Standard Model, so in first
approximation the experimental consequences should be small. However,
if our purpose is precisely to measure $g_R$ or at least to set a bound
on it, it is clear that the effect needs to be taken into account. As
already emphasized, to be able to tell left from right effective
couplings one absolutely needs to consider the spin of the top.

The next step is to select the direction
 where one is to `measure' the spin of the
top. By tracing the appropriate spin operator with the density
matrix one would determine the expected probabilities of finding a
top that (after the measure) would point in the given direction of
our choice. There is a privileged spin basis, namely the one where
the density matrix is diagonal, where the calculation is greatly
simplified since one needs not compute the off-diagonal terms.
This diagonalization process has to be done event by event and it
selects a particular vector $\hat{n}$ (event dependent). In
section \ref{diagonalization} we provide explicit formulae for
this privileged direction. Elementary Quantum Mechanical
considerations show that this is also the direction where the
differential cross section is maximal (or minimal depending on the
sign of the spin). Using this 3-vector as
spin basis, for instance, one can multiply the probability of
producing a top polarized in the positive $+\hat{n}$ direction
with the corresponding decay angular probability distribution plus
the probability of producing a top polarized in the negative
$-\hat{n}$ distribution times the the corresponding decay angular
probability distribution. The dependence on the effective left and
right couplings $g_L$ and $g_R$ is obviously contained in the
density matrix  and also in the decay distributions.

Obviously, since the entries of the density matrix depend
on the spin basis, the final physically observable result of
the previous analysis will certainly depend
on the spin basis too. How is this possible?
In fact this is as it should be;
we are multiplying probabilities
 and in fact we are neglecting
the quantum interference terms because we are assuming that the
polarization of the top is measured in the intermediate state
{\it before} the top quark decays.
Then there should be no surprise in that the interaction
between the top and the apparatus
measuring its spin  modifies the final
physical results.

However,
a proper measure of the top spin {\it before} it decays is impossible;
the only way we learn about top polarization is precisely
from the {\em final} decay products. So, the previous procedure
it is conceptually incorrect\footnote{Even if one is considering, as we
do here, only on-shell tops}. The final result has to be
strictly independent of the intermediate spin basis one uses.
Does this mean that the usual procedure ---which is the one we just
described--- is totally flawed? In principle yes, however
one expects that the coarsening involved
in the measuring process washes some or all of the interference effects.
Then perhaps
the previous procedure where one
assumes that the spin of the top is well defined  and one
proceeds as if it could be measured before it decays it
could be approximately correct.
But what are then the errors involved? Do they jeopardize the determination
of some of the effective couplings, in particular the distinction
between $g_L$ and $g_R$? These are some of the issues we would like to address
in the present work.

\section{The differential cross section for polarized top production}
\label{differentialc}

We shall discuss here the $t$-channel production for the sake
of definiteness. This is the most involved process. We refer
the reader to \cite{EM} for detailed expressions of the different
amplitudes.
We denote the matrix elements of the hard subprocess of
Fig. \ref{u+gt+b-d+tot} by $M_{+}^{d}$. There will also be a
$M_{+}^{\bar{u}}$, corresponding to having instead a $\bar{u}$ as spectator
quark. We will also eventually define the matrix elements
corresponding to the processes producing anti-tops as $M_{-}^{u}$, and $%
M_{-}^{\bar{d}}$. With these definitions the differential cross section for
polarized tops $d\sigma $ can be written schematically as
\begin{equation*}
d\sigma =\beta \left( f_{u}\left| M_{+}^{d}\right| ^{2}+f_{\bar{d}}\left|
M_{+}^{\bar{u}}\right| ^{2}\right) ,
\end{equation*}
where $f_{u}$ and $f_{\bar{d}}$ denote the parton distribution functions
corresponding to extracting a $u$-type quark and a $\bar{d}$-type quark
respectively and $\beta $ is a proportionality factor incorporating the
kinematics and also the gluon distribution function.
Using our analytical results for the matrix elements given
in the appendix of \cite{EM} we obtain for the differential cross-section

\begin{eqnarray}
d\sigma &=&\beta f_{u}\left[ \left| g_{L}\right| ^{2}\left( a+a_{n}\right)
+\left| g_{R}\right| ^{2}\left( b+b_{n}\right) +\frac{g_{R}^{\ast
}g_{L}+g_{R}g_{L}^{\ast }}{2}\left( c+c_{n}\right) +i\frac{g_{L}^{\ast
}g_{R}-g_{R}^{\ast }g_{L}}{2}d_{n}\right]  \notag \\
&+&\beta f_{\bar{d}}\left[ \left| g_{R}\right| ^{2}\left( a-a_{n}\right)
+\left| g_{L}\right| ^{2}\left( b-b_{n}\right) +\frac{g_{R}^{\ast
}g_{L}+g_{R}g_{L}^{\ast }}{2}\left( c-c_{n}\right) -i\frac{g_{L}^{\ast
}g_{R}-g_{R}^{\ast }g_{L}}{2}d_{n}\right] \notag  \\
&=&\left(
\begin{array}{cc}
g_{L}^{\ast } & g_{R}^{\ast }
\end{array}
\right) A\left(
\begin{array}{c}
g_{L} \\
g_{R}
\end{array}
\right) ,   \label{t-channeldecomp}
\end{eqnarray}
where
\begin{equation}
A=\beta \left(
\begin{array}{cc}
f_{u}\left( a+a_{n}\right) +f_{\bar{d}}\left( b-b_{n}\right) & \frac{1}{2}%
f_{u}\left( c+c_{n}+id_{n}\right) +\frac{1}{2}f_{\bar{d}}\left(
c-c_{n}-id_{n}\right) \\
\frac{1}{2}f_{u}\left( c+c_{n}-id_{n}\right) +\frac{1}{2}f_{\bar{d}}\left(
c-c_{n}+id_{n}\right) & f_{u}\left( b+b_{n}\right) +f_{\bar{d}}\left(
a-a_{n}\right)
\end{array}
\right) ,  \label{new}
\end{equation}
and where $a,$ $b$, $c$, $a_{n}$, $b_{n}$, $c_{n}$ and $d_{n}$ are
independent of the effective couplings $g_{R}$ and $g_{L}$ and the
subscripts $n$ indicate linear dependence on the top spin four-vector $n$.
All these quantities depend only on masses and momenta. The $c$,
$c_n$ and $d_n$ terms are proportional to the bottom mass and are
therefore absent if one neglects $m_b$ (this at first sight does
not look unreasonable, given the energies involved).
Inspection of the above differential cross-section reveals that
in the $m_b$ limit, the only way to tell left from right effective
couplings is precisely by considering and measuring
 polarized cross sections (the terms in $a_n$, $b_n$)
unless one is willing to rely strongly on the
parton distribution functions\footnote{The statement is exact if one uses
the so-called effective $W$ approximation, which is not terribly accurate
for the present case and certainly not recommended\cite{EM1}, but widely
used in LHC physics}.
For these reasons, both polarization and $m_b$ terms are quite important.

We observe that $A$ is an Hermitian matrix and
therefore it is diagonalizable with real eigenvalues. Moreover, from the
positivity of $d\sigma $ we immediately arrive at the constraints
\begin{eqnarray}
\det A &\geq &0,  \label{constr1} \\
TrA &\geq &0,  \label{constr2}
\end{eqnarray}
that is
\begin{eqnarray}
&&\left( f_{u}\left( a+a_{n}\right) +f_{\bar{d}}\left( b-b_{n}\right)
\right) \left( f_{u}\left( b+b_{n}\right) +f_{\bar{d}}\left( a-a_{n}\right)
\right)  \notag \\
&\geq &\frac{1}{4}\left( c^{2}\left( f_{u}+f_{\bar{d}}\right) ^{2}+\left(
c_{n}^{2}+d_{n}^{2}\right) \left( f_{u}-f_{\bar{d}}\right)
^{2}+2cc_{n}\left( f_{u}^{2}-f_{\bar{d}}^{2}\right) \right) ,  \label{co1}
\end{eqnarray}
and
\begin{equation}
\left( f_{u}+f_{\bar{d}}\right) \left( a+b\right) +\left( f_{u}-f_{\bar{d}%
}\right) \left( a_{n}+b_{n}\right) \geq 0.  \label{co2}
\end{equation}
Note that it is not possible to saturate both constraints for the same
configuration because this would imply a vanishing $A$ which in turn would
imply relations such as
\begin{equation*}
\frac{a+b}{a_{n}+b_{n}}=\frac{f_{\bar{d}}-f_{u}}{f_{\bar{d}}+f_{u}}=\frac{%
a_{n}-b_{n}}{a-b},
\end{equation*}
which evidently do not hold. Moreover, since constraints (\ref{co1}) and (%
\ref{co2}) must be satisfied for any set of positive parton
distribution functions we immediately obtain the bounds
\begin{eqnarray*}
ab+a_{n}b_{n}-\frac{1}{4}\left( c^{2}+c_{n}^{2}+d_{n}^{2}\right) &\geq
&\left| a_{n}b+ab_{n}-\frac{1}{2}cc_{n}\right| \\
b^{2}+a^{2}-\left( b_{n}^{2}+a_{n}^{2}\right) &\geq &\frac{1}{2}\left(
c^{2}-\left( c_{n}^{2}+d_{n}^{2}\right) \right) .
\end{eqnarray*}
In order to have a 100\% polarized top we need a spin four-vector $n$ that
saturates the constraint (\ref{constr1}) (that is Eq.(\ref{co1})) for each
kinematical situation, that is we need $A\left( n\right) $ to have a zero
eigenvalue which is equivalent to have a unitary matrix $C$ satisfying
\begin{equation*}
C^{\dagger }AC=\mathrm{diag}\left( \lambda ,0\right) ,
\end{equation*}
for some positive eigenvalue $\lambda $. In general such $n$ need not exist
and, should it exist, is in any case independent of the effective couplings $%
g_{R}$ and $g_{L}$. Moreover, provided this $n$ exists there is only one
solution (up to a global complex normalization factor $\alpha $) for the
pair $\left( g_{R},g_{L}\right) $ to the equation $d\sigma =0,$ This
solution is just
\begin{eqnarray}
g_{L} &=&\alpha C_{12},  \notag \\
g_{R} &=&\alpha C_{22}.  \label{100}
\end{eqnarray}
Note that if one of the effective couplings vanishes we can take the other
constant and arbitrary. However if both effective couplings are
non-vanishing we would have a quotient $g_{R}/g_{L}$ that would depend in
general on the kinematics. This is not possible so we can conclude that for
a non-vanishing $g_{R}$ ( $g_{L}$ is evidently non-vanishing) it is not
possible to have a pure spin state (or, else, only for fine tuned $g_{R}$ a
100\% polarization is possible).

Let us now give a very simple
 example to make the previous discussion more understandable:
in the un-physical situation where $m_{t}\rightarrow 0$ it can be
shown that there exists two solutions to the saturated constraint
(\ref{constr1}), namely
\begin{equation}
m_{t}n^{\mu }\rightarrow \pm \left( \left| \vec{p}_{1}\right| ,p_{1}^{0}%
\frac{\vec{p}_{1}}{\left| \vec{p}_{1}\right| }\right) ,
\end{equation}
once we have found this result we plug it in the expression (\ref{100}) and
we find the solutions $\left( 0,g_{L}\right) $ with $g_{L}$ arbitrary for
the $+$ sign and $\left( g_{R},0\right) $ with $g_{R}$ arbitrary for the $-$
sign. That is, physically we have zero probability of producing a right
handed top when we have only a left handed coupling and viceversa when we
have only a right handed coupling. Note that in this case it is clear that
having both effective couplings non-vanishing would imply the absence of 100
\% polarization in any spin basis. This can be understood in general
remembering that the top particle forms in general an entangled state with
the other particles of the process. Since we are tracing over the unknown
spin degrees of freedom and over the flavors of the spectator quark we do
not end up with a top in a pure polarized state.

\section{Cross-sections for top production and decay in the $s$-channel}
\label{xsection}

Let us  now turn to the $s$-channel process. This is, as already
mentioned, subdominant but non-negligible since it roughly amounts
to a 10\% of all single tops produced after our set of cuts are
imposed. It is also a lot cleaner from a theoretical point of
view, as QCD corrections are small.  As a by-product we shall
derive the differential decay width, which is applicable to both
the $t$ and $s$-channel processes.

Using the momenta conventions of Fig.
\ref{singletopschannelanddecay}
 and averaging over colors and spins of the initial fermions and summing over
colors and spins of the final fermions (remember that we have included a
spin projector for the top) the squared amplitude for top production is
given by
\begin{eqnarray}
\left| M_{n}\right| ^{2} &=&\frac{e^{4}N_{c}}{s_{W}^{4}}\left( \frac{1}{%
k^{2}-M_{W}^{2}}\right) ^{2}  \notag \\
&&\times \left\{ \left| \tilde{g}_{L}\right| ^{2}\left[ \left| g_{R}\right|
^{2}\left( q_{1}\cdot \frac{p_{1}+m_{t}n}{2}\right) \left( q_{2}\cdot \tilde{%
p}_{2}\right) +\left| g_{L}\right| ^{2}\left( q_{2}\cdot \frac{p_{1}-m_{t}n}{%
2}\right) \left( q_{1}\cdot \tilde{p}_{2}\right) \right. \right.  \notag \\
&&+m_{b}\frac{g_{L}g_{R}^{\ast }+g_{R}g_{L}^{\ast }}{4}\left[ m_{t}\left(
q_{1}\cdot q_{2}\right) +\left( q_{2}\cdot p_{1}\right) \left( q_{1}\cdot
n\right) -\left( q_{2}\cdot n\right) \left( q_{1}\cdot p_{1}\right) \right]
\notag \\
&&+\left. \left. im_{b}\frac{g_{L}g_{R}^{\ast }-g_{R}g_{L}^{\ast }}{4}%
\varepsilon _{\mu \alpha \rho \sigma }n^{\mu }p_{1}^{\alpha }q_{1}^{\rho
}q_{2}^{\sigma }\right] \right\}  \notag \\
&&+\left| \tilde{g}_{R}\right| ^{2}\left[ \left| g_{R}\right| ^{2}\left(
q_{2}\cdot \frac{p_{1}+m_{t}n}{2}\right) \left( q_{1}\cdot \tilde{p}%
_{2}\right) +\left| g_{L}\right| ^{2}\left( q_{1}\cdot \frac{p_{1}-m_{t}n}{2}%
\right) \left( q_{2}\cdot \tilde{p}_{2}\right) \right.  \notag \\
&&+m_{b}\frac{g_{L}g_{R}^{\ast }+g_{R}g_{L}^{\ast }}{4}\left[ m_{t}\left(
q_{1}\cdot q_{2}\right) +\left( q_{1}\cdot p_{1}\right) \left( q_{2}\cdot
n\right) -\left( q_{1}\cdot n\right) \left( q_{2}\cdot p_{1}\right) \right]
\notag \\
&&+\left. \left. im_{b}\frac{g_{L}g_{R}^{\ast }-g_{R}g_{L}^{\ast }}{4}%
\varepsilon _{\mu \alpha \rho \sigma }n^{\mu }p_{1}^{\alpha }q_{2}^{\rho
}q_{1}^{\sigma }\right] \right\} ,  \label{schanproduction}
\end{eqnarray}
where $\tilde{g}_{L},$ and $\tilde{g}_{R}$ are left and right couplings to
light quarks and $g_{L}$ and $g_{R}$ are the effective
couplings to the top- bottom
system. In the numerical results we have taken $\tilde{g}_{L}=1,$
$\tilde{g}%
_{R}=0$; i.e. we stick to the tree-level Standard Model values
in the light sector, but is quite straightforward to include
more general couplings.  Notice that, exactly as for the $t$-channel,
the crossed $g_Lg_R$ terms vanish in the differential cross-section
in the $m_b\to 0$ limit. Also for exactly
the same reasons as in the $t$-channel analysis, modulo parton
distribution functions effects, the differential unpolarized
production cross section would be proportional to
$\vert g_L\vert^2 + \vert g_R\vert^2$.

The differential cross section for producing polarized tops
is then
\begin{eqnarray*}
d\sigma _{\hat{n}} &=&f\left( \tilde{x}_{1},\tilde{x}_{2},\left(
q_{1}+q_{2}\right) ^{2},\Lambda _{QCD}\right) d\tilde{x}_{1}d\tilde{x}_{2}%
\frac{1}{4\left| q_{2}^{0}\overrightarrow{q_{1}}-\overrightarrow{q_{2}}%
q_{1}^{0}\right| } \\
&&\times \frac{d^{3}p_{1}}{\left( 2\pi \right) ^{3}2p_{1}^{0}}\frac{d^{3}%
\tilde{p}_{2}}{\left( 2\pi \right) ^{3}2\tilde{p}_{2}^{0}}\left|
M_{n}\right| ^{2}\left( 2\pi \right) ^{4}\delta ^{4}\left(
q_{1}+q_{2}-p_{1}-p_{2}\right)
\end{eqnarray*}
where $f\left( \tilde{x}_{1},\tilde{x}_{2},\left( q_{1}+q_{2}\right)
^{2},\Lambda _{QCD}\right) d\tilde{x}_{1}d\tilde{x}_{2}$ accounts for the
quarks parton distribution functions.

The total decay rate of the top, on the other hand, with arbitrary left and
right effective couplings is given by
\begin{eqnarray*}
\Gamma &=&\frac{e^{2}}{s_{W}^{2}}\left\{ \left( \left| g_{L}\right|
^{2}+\left| g_{R}\right| ^{2}\right) \left( m_{t}^{2}+m_{b}^{2}-2M_{W}^{2}+%
\frac{\left( m_{t}^{2}-m_{b}^{2}\right) ^{2}}{M_{W}^{2}}\right) \right. \\
&&\left. -12m_{t}m_{b}\frac{g_{L}g_{R}^{\ast }+g_{R}g_{L}^{\ast }}{2}%
\right\} \frac{\sqrt{\left( m_{t}^{2}+m_{b}^{2}-M_{W}^{2}\right)
^{2}-4m_{t}^{2}m_{b}^{2}}}{64\pi m_{t}^{2}p_{1}^{0}}.
\end{eqnarray*}
The squared amplitude corresponding to the decay rate in the channel
depicted in Fig. (\ref{singletopschannelanddecay}) summing over the top
polarizations (with a spin projector inserted), averaging over its color
and summing over colors and polarizations of decay products is given by
\begin{equation}
\left| M_{n}^{D}\right| ^{2}=-\frac{4}{N_{c}}\left| M_{n}\right| ^{2}\left(
q_{1}\rightarrow k_{2},\ q_{2}\rightarrow k_{1},\ \tilde{p}_{2}\rightarrow
-p_{2}\right) , \label{xurro}
\end{equation}
where $\left| M_{n}\right| ^{2}\left( q_{1}\rightarrow k_{2},\
q_{2}\rightarrow k_{1},\ \tilde{p}_{2}\rightarrow -p_{2}\right) $ is just
expression (\ref{schanproduction}) with the indicated changes in momenta.
In the above expression
$\tilde{g}_{L},$ and $\tilde{g}_{R}$ are the left and right couplings
corresponding to the lepton-neutrino vertex. We have assumed $\tilde{g%
}_{L}=1,$ $\tilde{g}_{R}=0$, but again this hypothesis can be relaxed.
The decay rate differential distribution
for this channel is given by
\begin{equation*}
d\Gamma _{n}=\frac{\left| M_{n}^{D}\right| ^{2}}{2p_{1}^{0}}\frac{d^{3}k_{1}%
}{\left( 2\pi \right) ^{3}2k_{2}^{0}}\frac{d^{3}k_{2}}{\left( 2\pi \right)
^{3}2k_{1}^{0}}\frac{d^{3}p_{2}}{\left( 2\pi \right) ^{3}2p_{2}^{0}}\left(
2\pi \right) ^{4}\delta ^{4}\left( k_{1}+k_{2}+p_{2}-p_{1}\right) .
\end{equation*}
Finally, using the narrow-width approximation, we have that the
differential cross section $d\sigma $ corresponding to Fig.
\ref{singletopschannelanddecay} is given by
\begin{equation}
d\sigma =\sum_{\pm n}d\sigma _{n}\times \frac{d\Gamma _{n}}{\Gamma }.
\label{finalschannel}
\end{equation}

\section{The role of spin in the narrow-width approximation}
\label{spinrole}

Within the narrow-width approximation we just discussed we
decompose the process depicted in Fig.
\ref{singletopschannelanddecay} in two consecutive processes: the
top production and its consecutive decay. In that set up we denote
the single top production amplitude as $A_{p,\pm \hat{n}\left(
p\right) }$ and the top decay amplitude as $B_{p,\pm \hat{n}\left(
p\right) }.$ In the polar representation we write
\begin{eqnarray*}
A_{p,\pm \hat{n}\left( p\right) } &=&\left| A_{p,\pm \hat{n}\left( p\right)
}\right| e^{i\varphi _{\pm }\left( p\right) }, \\
B_{p,\pm \hat{n}\left( p\right) } &=&\left| B_{p,\pm \hat{n}\left( p\right)
}\right| e^{i\omega _{\pm }\left( p\right) },
\end{eqnarray*}
where $p$ indicate external momenta and $\hat{n}\left( p\right) $ a given
spin basis for the top. The differential cross section for the whole process
is schematically given by
\begin{equation}
d\mathcal{\sigma }=\int \left| A_{p,+\hat{n}\left( p\right) }B_{p,+\hat{n}%
\left( p\right) }+A_{p,-\hat{n}\left( p\right) }B_{p,-\hat{n}\left( p\right)
}\right| ^{2}dp.  \label{exact}
\end{equation}
Hence
\begin{eqnarray}
d\mathcal{\sigma } &=&\int \left| A_{p,+\hat{n}\left( p\right) }\right|
^{2}\left| B_{p,+\hat{n}\left( p\right) }\right| ^{2}dp+\int \left| A_{p,-%
\hat{n}\left( p\right) }\right| ^{2}\left| B_{p,-\hat{n}\left( p\right)
}\right| ^{2}dp  \notag \\
&&+2\int \left| A_{p,+\hat{n}\left( p\right) }\right| \left| B_{p,+\hat{n}%
\left( p\right) }\right| \left| A_{p,-\hat{n}\left( p\right) }\right| \left|
B_{p,-\hat{n}\left( p\right) }\right|  \notag \\
&&\times \cos \left( \varphi _{+}\left( p\right) -\varphi _{-}\left(
p\right) +\omega _{+}\left( p\right) -\omega _{-}\left( p\right) \right) dp
\notag \\
&\simeq &\int \left| A_{p,+\hat{n}\left( p\right) }\right| ^{2}\left| B_{p,+%
\hat{n}\left( p\right) }\right| ^{2}dp+\int \left| A_{p,-\hat{n}\left(
p\right) }\right| ^{2}\left| B_{p,-\hat{n}\left( p\right) }\right| ^{2}dp.
\label{approx}
\end{eqnarray}
Since the axis with respect to which the spin basis is defined is
completely arbitrary $d\mathcal{\sigma }$ is independent on this
choice of basis. However within the narrow width approximation one
never computes $d\mathcal{\sigma }$ following formula
(\ref{exact}). The commonly used procedure \cite{mandp,Mauser}
consists in computing the probability of producing a polarized top
and then multiplying this probability by the probability of a
given decay channel (see Eq. (\ref{finalschannel})). This
procedure is equivalent to the neglection of the interference term
in formula (\ref{approx}) as indicated there. First of all, as
discussed in the introduction, if one neglects the interference
term, the result depends on the spin basis; i.e. on the direction
one chooses to measure the third component of the top spin. This
is of course acceptable if one really performs a physical measure
of the spin (in the $\hat{n(p)}$ direction in this case) since the
interaction with the apparatus modifies the state. A dependence on
the spin frame is however
 unacceptable if the spin is not measured {\it
before} the top decays.

Let us see whether this approximation can justified nevertheless.
Clearly, the
integration over momenta enhances the positive-definite terms in front of
the interference oscillating one. If in addition we make a choice for
$\hat{n
}\left( p\right) $ that diagonalizes the top spin density matrix
 and thus maximizes $\left| A_{p,+\hat{n}
\left( p\right) }\right| $ and minimizes $\left| A_{p,-\hat{n}\left(
p\right) }\right| $, then we expect the interference term to be negligible
when compared to $\int \left| A_{p,+\hat{n}\left( p\right) }\right|
^{2}\left| B_{p,+\hat{n}\left( p\right) }\right| ^{2}dp$ even for small
amount of phase space integration. In the \textit{s}-channel we will see in
the next section that the limit of $g_{R}\rightarrow 0$ there exists a spin
basis $\hat{n}\left( p\right) $ where $\left| A_{p,-\hat{n}\left( p\right)
}\right| $ is strictly zero. This basis is given by
\begin{equation*}
n=\frac{1}{m_{t}}\left( \frac{m_{t}^{2}}{\left( q_{2}\cdot p_{1}\right) }%
q_{2}-p_{1}\right) .
\end{equation*}
From this it follows that for small $g_{R}$ if we use that basis the
interference integrand is already negligible with respect to the dominant
term $\int \left| A_{p,+\hat{n}\left( p\right) }\right| ^{2}\left| B_{p,+%
\hat{n}\left( p\right) }\right| ^{2}dp$. For $g_{R}\neq 0$ one can still
find a basis that maximizes $\left| A_{p,+\hat{n}\left( p\right) }\right| $
(and minimizes $\left| A_{p,-\hat{n}\left( p\right) }\right| $) and
therefore diagonalizes the top density matrix $\rho $.
In the next section we will show
how to obtain such a basis that will be the one used in our numerical
integration. In these simulations we have checked numerically that this
basis is the one that maximizes $d\mathcal{\sigma }$ and therefore, on the
same grounds, the one that minimizes the interference term.
The same considerations can be applied to the $t$-channel process.

Given that the observables are strictly independent of the choice of spin
basis \emph{only} if the interference term is included, we can easily assess
the importance of the latter by checking to what extent a residual spin
basis dependence is present. We have checked numerically this point by
changing the definition of the spin basis $\hat{n}\left( p\right) $ and
noting that our results are actually only weakly dependent on the
choice of $\hat{n}\left(
p\right) $ even for a small amount of coarsening.
A 4\% maximum variation in $p_T$ distributions
was found between the optimal diagonal basis
and another basis orthogonal to the beam axis
(that is, almost orthogonal to all
momenta). Moreover we have checked that if spin is ignored altogether
(by considering unpolarized top production) roughly the
same amount of variation with respect to the diagonal basis is
observed. Thus we conclude that even though the
dependence on the choice of spin basis is not dramatic, its consideration is
a must for a precise description using the narrow-width approximation taking
into account the presumed smallness of the effective coupling to be
measured and how subtle the experimental distinction of left and right
couplings turns out to be.

\section{The diagonal basis}
\label{diagonalization}

As stated in the previous section in order to calculate the top decay we
have to find the basis where the polarized single top production cross
section is maximal. The can do this maximizing in the $4$-dimensional space
generated by the components of $n$ constrained by
\begin{equation}
n\cdot p_{1}=0,\qquad n^{2}=-1,  \label{spinconstraints}
\end{equation}
where $p_{1}$ is the top four-moment, that is
\begin{eqnarray*}
n^{0} &=&\frac{n^{1}p_{1}^{1}+n^{2}p_{1}^{2}+n^{2}p_{1}^{2}}{p_{1}^{0}}, \\
\left( p_{1}^{0}\right) ^{2} &=&\left( p_{1}^{0}\right) ^{2}\left\| \vec{n}%
\right\| ^{2}-\left( n^{1}p_{1}^{1}+n^{2}p_{1}^{2}+n^{2}p_{1}^{2}\right)
^{2},
\end{eqnarray*}
where $\left\| \vec{n}\right\| =\sqrt{\left( n^{1}\right) ^{2}+\left(
n^{2}\right) ^{2}+\left( n^{3}\right) ^{2}}$, that is $n^{i}=\left\| \vec{n}%
\right\| \hat{n}^{i}$ with $\hat{n}$ the normalized spin three-vector. From
above equations we obtain
\begin{eqnarray*}
\left\| \vec{n}\right\| &=&\frac{p_{1}^{0}}{\sqrt{\left( p_{1}^{0}\right)
^{2}-\left( \hat{n}^{1}p_{1}^{1}+\hat{n}^{2}p_{1}^{2}+\hat{n}%
^{2}p_{1}^{2}\right) ^{2}}}, \\
n^{0} &=&\left\| \vec{n}\right\| \frac{\hat{n}^{1}p_{1}^{1}+\hat{n}%
^{2}p_{1}^{2}+\hat{n}^{2}p_{1}^{2}}{p_{1}^{0}},
\end{eqnarray*}
from which Eq. (\ref{spinfour}) follows immediately. Let us now find the
polarization vector that maximizes and minimizes the differential cross
section of single top production.

\subsection{The \textit{t}-channel}

We will begin with the \textit{t}-channel the was analyzed in the previous
chapter. Using Eq. (\ref{t-channeldecomp}) we define
\begin{eqnarray}
a_{n} &=&n\cdot a,\qquad b_{n}=n\cdot b,  \notag \\
c_{n} &=&n\cdot c,\qquad d_{n}=n\cdot d,  \label{ndecomp}
\end{eqnarray}
and using Lagrange multipliers $\lambda _{1}$ and $\lambda _{2}$ for
constraints (\ref{spinconstraints}) we maximize
\begin{equation*}
\sigma +\lambda _{1}\left( n^{2}+1\right) +\lambda _{2}n\cdot p_{1},
\end{equation*}
obtaining the equations
\begin{eqnarray}
n &=&-\frac{\beta }{2\lambda _{1}}f_{u}\left[ \left| g_{L}\right|
^{2}a+\left| g_{R}\right| ^{2}b+\frac{g_{R}^{\ast }g_{L}+g_{R}g_{L}^{\ast }}{%
2}c+i\frac{g_{L}^{\ast }g_{R}-g_{R}^{\ast }g_{L}}{2}d\right]  \notag \\
&&+\frac{\beta }{2\lambda _{1}}f_{\bar{d}}\left[ \left| g_{R}\right|
^{2}a+\left| g_{L}\right| ^{2}b+\frac{g_{R}^{\ast }g_{L}+g_{R}g_{L}^{\ast }}{%
2}c+i\frac{g_{L}^{\ast }g_{R}-g_{R}^{\ast }g_{L}}{2}d\right] -\frac{\lambda
_{2}}{2\lambda _{1}}p_{1},  \label{seq1} \\
0 &=&n^{2}+1,  \label{seq2} \\
0 &=&n\cdot p_{1},  \label{seq3}
\end{eqnarray}
and thus using Eqs. (\ref{seq1}) and (\ref{seq3})
\begin{eqnarray}
\lambda _{2} &=&-\frac{\beta }{m_{t}^{2}}f_{u}\left[ \left| g_{L}\right|
^{2}a\cdot p_{1}+\left| g_{R}\right| ^{2}b\cdot p_{1}+\frac{g_{R}^{\ast
}g_{L}+g_{R}g_{L}^{\ast }}{2}c\cdot p_{1}+i\frac{g_{L}^{\ast
}g_{R}-g_{R}^{\ast }g_{L}}{2}d\cdot p_{1}\right]  \notag \\
&&+\frac{\beta }{m_{t}^{2}}f_{\bar{d}}\left[ \left| g_{R}\right| ^{2}a\cdot
p_{1}+\left| g_{L}\right| ^{2}b\cdot p_{1}+\frac{g_{R}^{\ast
}g_{L}+g_{R}g_{L}^{\ast }}{2}c\cdot p_{1}+i\frac{g_{L}^{\ast
}g_{R}-g_{R}^{\ast }g_{L}}{2}d\cdot p_{1}\right] ,  \notag
\end{eqnarray}
and therefore
\begin{eqnarray*}
n &=&\frac{\beta }{2\lambda _{1}}\left\{ \left( f_{u}\left| g_{L}\right|
^{2}-f_{\bar{d}}\left| g_{R}\right| ^{2}\right) \left( \frac{a\cdot p_{1}}{%
m_{t}^{2}}p_{1}-a\right) +\left( f_{u}\left| g_{R}\right| ^{2}-f_{\bar{d}%
}\left| g_{L}\right| ^{2}\right) \left( \frac{b\cdot p_{1}}{m_{t}^{2}}%
p_{1}-b\right) \right. \\
&&+\left. \frac{g_{R}^{\ast }g_{L}+g_{R}g_{L}^{\ast }}{2}\left( f_{u}-f_{%
\bar{d}}\right) \left( \frac{c\cdot p_{1}}{m_{t}^{2}}p_{1}-c\right) +i\frac{%
g_{L}^{\ast }g_{R}-g_{R}^{\ast }g_{L}}{2}\left( f_{u}-f_{\bar{d}}\right)
\left( \frac{d\cdot p_{1}}{m_{t}^{2}}p_{1}-d\right) \right\} ,
\end{eqnarray*}
with the normalization factor $\lambda _{1}$ given by Eq. (\ref{seq2}). Note
that in the idealized case $f_{u}=f_{\bar{d}}=f$ we obtain
\begin{equation*}
n=\alpha \left\{ \frac{\left( a-b\right) \cdot p_{1}}{m_{t}^{2}}p_{1}-\left(
a-b\right) \right\} ,
\end{equation*}
where $\alpha $ is the normalization constant that does not depend on $f$ or
the effective couplings. In the SM ($g_{R}=0$) we obtain
\begin{equation*}
n=\alpha \left( f_{u}\left( \frac{a\cdot p_{1}}{m_{t}^{2}}p_{1}-a\right) -f_{%
\bar{d}}\left( \frac{b\cdot p_{1}}{m_{t}^{2}}p_{1}-b\right) \right) ,
\end{equation*}
where $\alpha $ is a normalizing factor.

\subsection{The \textit{s}-channel}

The \textit{s}-channel differential cross section has the form
\begin{eqnarray*}
d\sigma &=&\beta \left( f_{u}f_{\bar{d}}+f_{c}f_{\bar{s}}\right) \left[
\left| g_{L}\right| ^{2}\left( a_{s}+a_{n}\right) +\left| g_{R}\right|
^{2}\left( b_{s}+b_{n}\right) \right. \\
&&+\left. \frac{g_{R}^{\ast }g_{L}+g_{R}g_{L}^{\ast }}{2}\left(
c_{s}+c_{n}\right) +i\frac{g_{L}^{\ast }g_{R}-g_{R}^{\ast }g_{L}}{2}d_{n}%
\right] ,
\end{eqnarray*}
where again $\beta $ is a proportionality incorporating the kinematics, and
where $f_{u,c}$ and $f_{\bar{d},\bar{s}}$ denote the parton distribution
functions corresponding to extracting a $u,c$-type quarks and a $\bar{d},%
\bar{s}$-type quarks respectively. Using again the decomposition (\ref
{ndecomp}) and proceeding analogously to the \textit{t}-channel calculation
we obtain
\begin{eqnarray}
n &=&\alpha \left\{ \left| g_{L}\right| ^{2}\left( \frac{a\cdot p_{1}}{%
m_{t}^{2}}p_{1}-a\right) +\left| g_{R}\right| ^{2}\left( \frac{b\cdot p_{1}}{%
m_{t}^{2}}p_{1}-b\right) \right.  \notag \\
&&+\left. \frac{g_{R}^{\ast }g_{L}+g_{R}g_{L}^{\ast }}{2}\left( \frac{c\cdot
p_{1}}{m_{t}^{2}}p_{1}-c\right) +i\frac{g_{L}^{\ast }g_{R}-g_{R}^{\ast }g_{L}%
}{2}\left( \frac{d\cdot p_{1}}{m_{t}^{2}}p_{1}-d\right) \right\} ,
\label{nschan}
\end{eqnarray}
where $\alpha $ is the normalizing factor that in this case (unlike in the
\textit{t}-channel result) does not depend on the parton distribution
 functions. From Eq. (\ref
{schanproduction}) we obtain
\begin{eqnarray*}
a^{\mu } &=&-m_{t}q_{2}^{\mu }\left( q_{1}\cdot \tilde{p}_{2}\right) , \\
b^{\mu } &=&+m_{t}q_{1}^{\mu }\left( q_{2}\cdot \tilde{p}_{2}\right) , \\
c^{\mu } &=&+m_{b}\left( q_{1}^{\mu }\left( q_{2}\cdot p_{1}\right)
-q_{2}^{\mu }\left( q_{1}\cdot p_{1}\right) \right) , \\
d^{\mu } &=&-m_{b}\varepsilon _{~\alpha \rho \sigma }^{\mu }p_{1}^{\alpha
}q_{1}^{\rho }q_{2}^{\sigma },
\end{eqnarray*}
hence replacing in Eq. (\ref{nschan}) we arrive at
\begin{eqnarray}
n^{\mu } &=&\alpha \left\{ \left| g_{L}\right| ^{2}\left( \left( q_{1}\cdot
\tilde{p}_{2}\right) \left( q_{2}\cdot p_{1}\right) p_{1}^{\mu }-\left(
q_{1}\cdot \tilde{p}_{2}\right) m_{t}^{2}q_{2}^{\mu }\right) \right.  \notag
\\
&&+\left| g_{R}\right| ^{2}\left( \left( q_{2}\cdot \tilde{p}_{2}\right)
\left( q_{1}\cdot p_{1}\right) p_{1}^{\mu }-\left( q_{2}\cdot \tilde{p}%
_{2}\right) m_{t}^{2}q_{1}^{\mu }\right)  \notag \\
&&+m_{b}m_{t}\frac{g_{R}^{\ast }g_{L}+g_{R}g_{L}^{\ast }}{2}\left(
q_{1}^{\mu }\left( q_{2}\cdot p_{1}\right) -q_{2}^{\mu }\left( q_{1}\cdot
p_{1}\right) \right)  \notag \\
&&+\left. i\frac{g_{R}^{\ast }g_{L}-g_{L}^{\ast }g_{R}}{2}%
m_{b}m_{t}\varepsilon _{~~\alpha \rho \sigma }^{\mu }p_{1}^{\alpha
}q_{1}^{\rho }q_{2}^{\sigma }\right\} ,  \label{finalbasis}
\end{eqnarray}
which is the basis we use in our numerical simulations. If we neglect $g_{R}$
we obtain
\begin{equation*}
n^{\mu }=\pm \frac{\left( q_{1}\cdot \tilde{p}_{2}\right) \left( q_{2}\cdot
p_{1}\right) p_{1}^{\mu }-\left( q_{1}\cdot \tilde{p}_{2}\right)
m_{t}^{2}q_{2}^{\mu }}{\sqrt{\left( q_{1}\cdot \tilde{p}_{2}\right)
^{2}\left( q_{2}\cdot p_{1}\right) ^{2}m_{t}^{2}-\left( q_{1}\cdot \tilde{p}%
_{2}\right) ^{2}m_{t}^{4}q_{2}^{2}}},
\end{equation*}
where we have included the normalization factor and since $q_{2}^{2}=0$ the
above reduces to
\begin{equation*}
m_{t}n=\pm \left( \frac{m_{t}^{2}}{\left( q_{2}\cdot p_{1}\right) }%
q_{2}-p_{1}\right) ,
\end{equation*}
which is the result we have quoted in the previous section coinciding with
\cite{mandp}

\section{Numerical analysis of $s$-channel single top production}

Let us start this section by
discussing the experimental cuts we have implemented. Due to geometrical
detector constraints\cite{atlas} we cut off very low angles
for the outgoing particles.
The charged particles in the final state have to come out with an angle in
between 10 and 170 degrees to be detected.
These angular cuts correspond to a cut in
pseudorapidity $\left| \eta \right| <2.44$. In order to be able to separate
the jets corresponding to the outgoing particles we implement
isolation cuts of 20 degrees between each other. These are the appropriate
cuts for general purpose experiments such as ATLAS or CMS.

The set of cuts used in this work are compatible with the ones used in
the \textit{t}-channel. Since in the previous paper \cite{EM}
top decay was not
considered, the equivalence is only approximate and a more detailed
phenomenological analysis will be required in due course (it is actually
quite straightforward with the help of the results presented here to redo
the $t$-channel study, but this goes beyond the scope of this paper).
The present analysis
should however suffice in any case to identify the most promising observables
and get a rough estimate of the precision that it can be reached.

We use a lower cut of 20 GeV in the $\bar{b}$ jet\footnote{In the
previous paper \cite{EM} the value used was 30GeV. We have decided
to use this lower value here to have a larger total cross-section
without compromising the theoretical accuracy}.  This
completely eliminates top production from a $b$-quark from the proton
sea and greatly reduces higher order QCD contributions. In the $t$-channel
reduces the cross section to about one third of its total value, since
typically the $\bar{b}$ quark comes out in the same direction as the
incoming gluon and a large fraction of them do not pass the cut.
Similarly, $p_{T}>20$ GeV cuts are
set for the top and spectator quark jets. These cuts guarantee the validity
of perturbation theory and will serve to separate from the overwhelming
background of low $p_{T}$ physics. These values come as a compromise to
preserve a good signal, while suppressing unwanted contributions. They are
very similar to the ones used in \cite{SSW} and \cite{mandp}.

In order to calculate the cross section $\sigma $ of the
process $pp\rightarrow t\bar{b}$ we have used the CTEQ4 set of structure
functions \cite{CTEQ4} to determine the probability of extracting a parton
with a given fraction of momenta from the proton.
To calculate the total event production corresponding to different observables
we have used the integrating Monte Carlo program VEGAS \cite{vegas}. We
present results after one year (defined as 10$^{7}$ seg.) run at full
luminosity in one detector (100 $\mathrm{fb}^{-1}$ at LHC).

Since in order to be able to perform the effective $Wt\bar{b}$ coupling
one definitely needs to tag the two $b$-type quarks, this value
for the luminosity is surely too high. The $b$ physics program
at ATLAS \cite{lluisa}, for instance, it is
planned to be done at one tenth of the total luminosity to avoid
pile-up effects. This is even more so in a dedicated
detector
such as LHCb\footnote{This type of analysis is anyway
not well suited for such a detector. The rapidity
for LHCb is in the range $1.6 < \eta < 4.9$ and the
angular separation cut between jets imposed here is
totally unfeasible. Furthermore, jet reconstruction
is not possible. Clearly the implementation
of this type of physics to this detector requires
a lot more ingenuity.}\cite{Amato:1998xt}
where the appropriate figure is expected
to be 2 fb$^{-1}$. ATLAS plans to do most of the $b$-physics runs before
full luminosity is reached, for instance. We have nonetheless
used the high luminosity figure since at this stage the
experimental strategy
is not totally settled yet.

The way we proceed is the following. We analyze the kinematics of
each event including a $\bar{b}$ that passes the experimental cuts
and reconstruct the vector $\hat{n}$ using the analytic formulae
presented in the previous sections. As the reader will remember,
this provides us with a spin basis that minimizes the quantum
interference terms. We then proceed to multiply the probabilities
classically ---just as if we pretend that the top spin has been
measured in the direction determined by $\hat{n}$ and we determine
the decay probability distribution. We retain only those final
states that pass the remaining cuts.

In the same way  and choosing arbitrary spin directions we are
able to see how much the physical results depend on the interference
term. We have found a 4\% difference between the worst case (a spin
direction perpendicular to almost all 3-momenta involved) and
the optimal case (found analytically here). We have every reason
to believe that, after the integration over momenta and the resulting
coarsening, the interference term is in this basis all but negligible.
The rest of the results presented in this section are all worked
out in the optimal spin basis.

\begin{figure}[!hbp]
\begin{center}
\includegraphics[width=130mm]{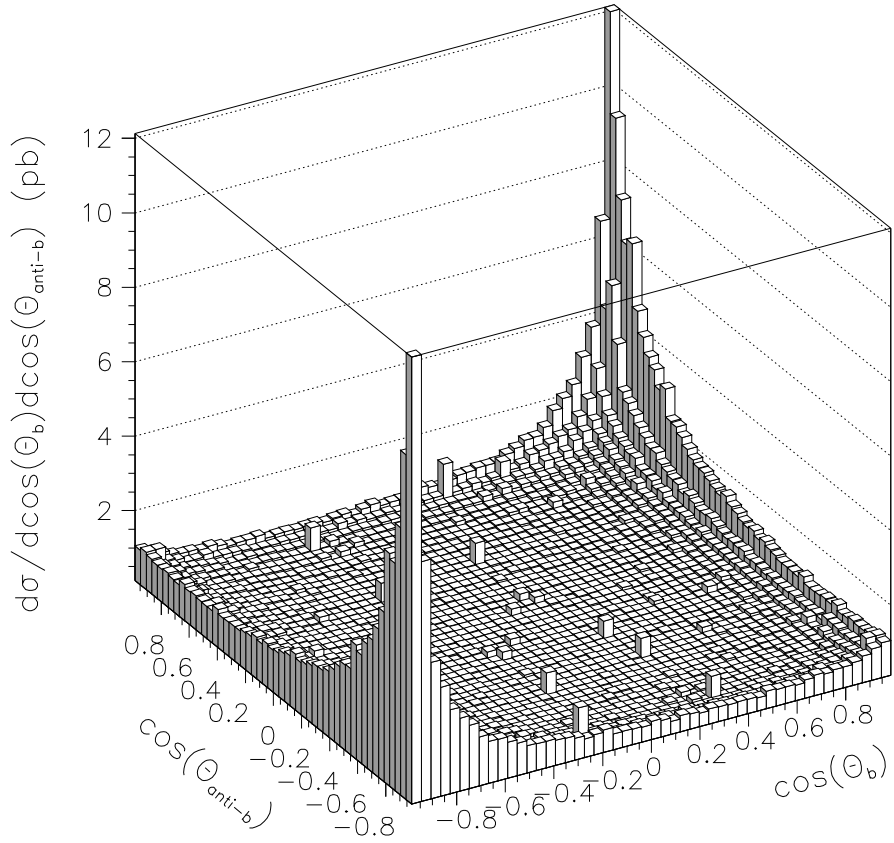}
\end{center}
\par
\caption{Distribution of the cosines of the polar angles of the
bottom and anti-bottom with respect to the beam line. The plot
corresponds to single top production at the LHC with top decay
included. The calculation was performed at the tree level in
Standard Model. For the parton
distribution functions we use $\protect\mu ^{2}=\hat{s}%
=\left( q_{1}+q_{2}\right) ^{2}$.}
\label{cosbcosab_gl=1_gr=0}
\end{figure}

\begin{figure}[!hbp]
\begin{center}
\includegraphics[width=80mm]{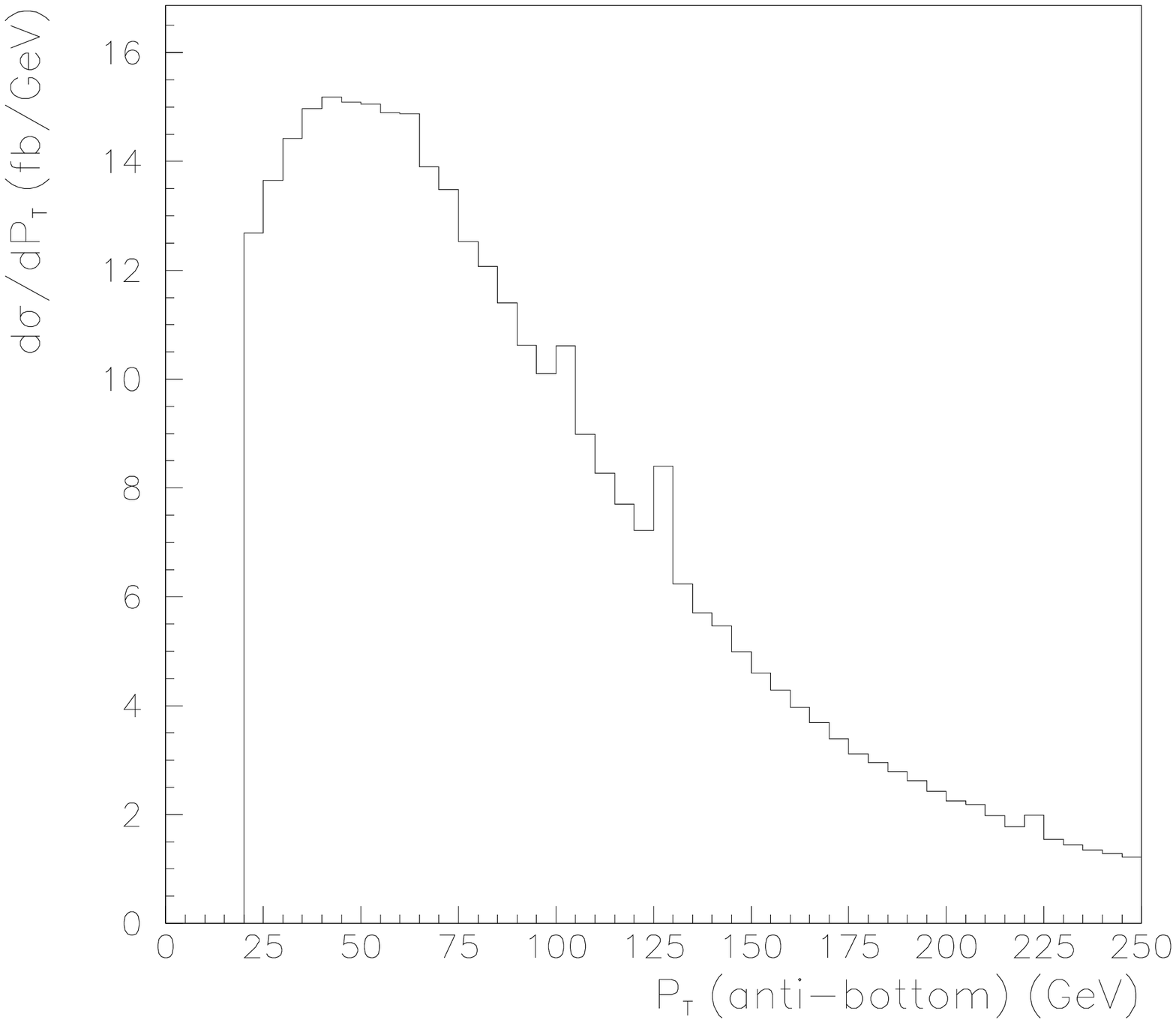}
\end{center}
\par
\caption{Anti-bottom transversal momentum distribution corresponding to
single top production at the LHC. The calculation has been performed at tree
level in the SM ($g_{L}=1$, $g_{R}=0$). }
\label{pabtrans_gl=1_gr=0}
\end{figure}

\begin{figure}[!hbp]
\begin{center}
\includegraphics[width=8cm]{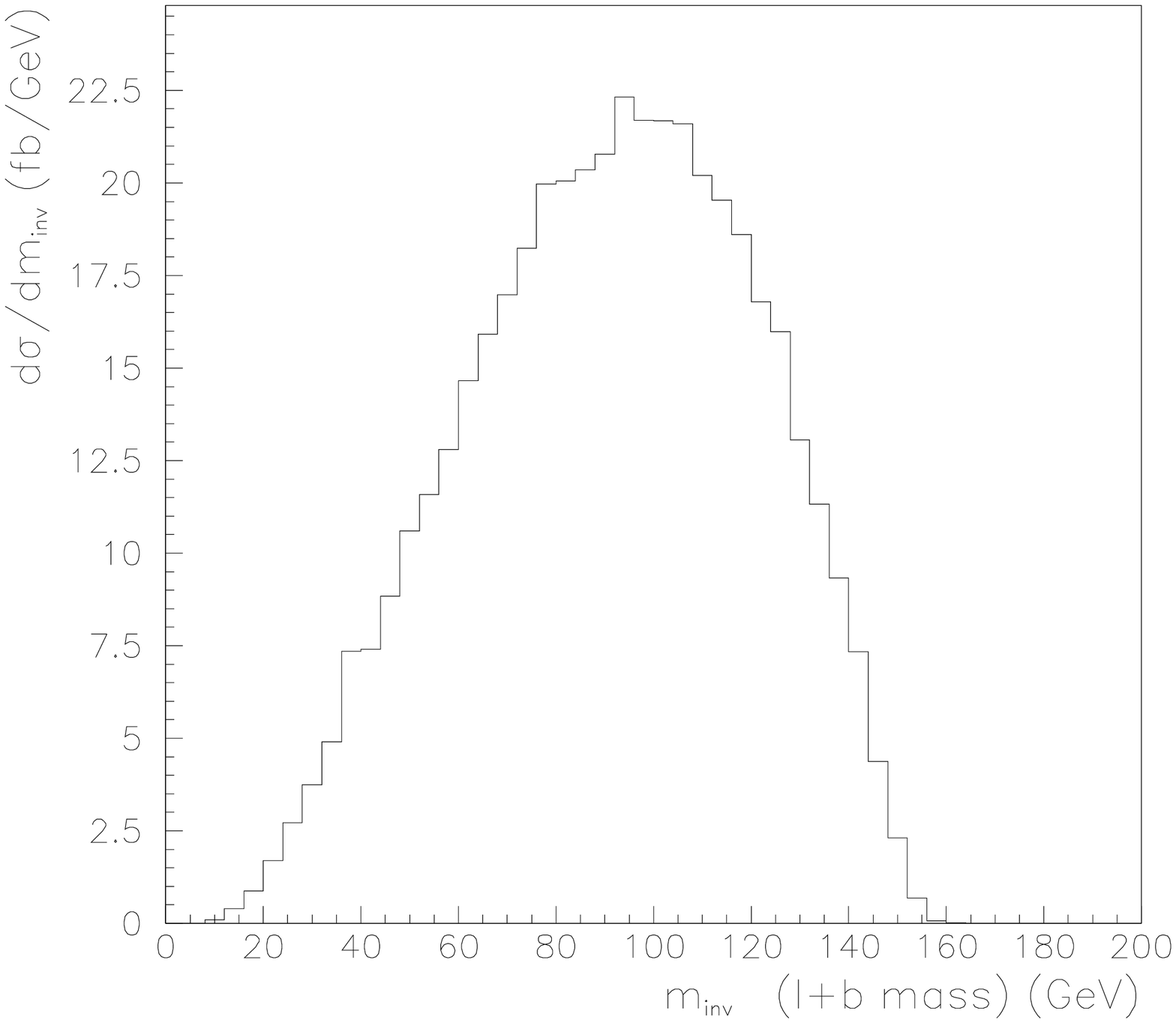}
\end{center}
\par
\caption{Distribution of the invariant mass of the lepton (electron or muon)
plus bottom system arising in top decay from single top production at the
LHC. The calculation was performed at the tree level in Standard Model with $%
\protect\mu ^{2}=\hat{s}=\left( q_{1}+q_{2}\right) ^{2}$.}
\label{plpbmvv_gl=1_gr=0}
\end{figure}

Let us first review the results obtained in the framework of the
tree-level standard model.
This corresponds to taking $g_L=1$ and $g_R=0$ in all our formulae.
 The results are summarized in Figs.
\ref{cosbcosab_gl=1_gr=0}, \ref{pabtrans_gl=1_gr=0},
\ref{plpbmvv_gl=1_gr=0}. The first of these figures show how
the final products of the process are predominantly emitted in
the axis direction (albeit not so much as in the case of the $t$-channel
production) and in the same direction. The plot shows the direction
respect to the beam of bottom and anti-bottom. Recall that a 10 degree
cut is implemented, as well as a separation cut of 20 degrees
among all jets. Fig. \ref{pabtrans_gl=1_gr=0}
shows the $p_T$ distribution for the $\bar{b}$, showing the 20
GeV cut on the $p_T$ of the $\bar{b}$ enforced.

\begin{figure}[!hbp]
\begin{center}
\includegraphics[width=80mm]{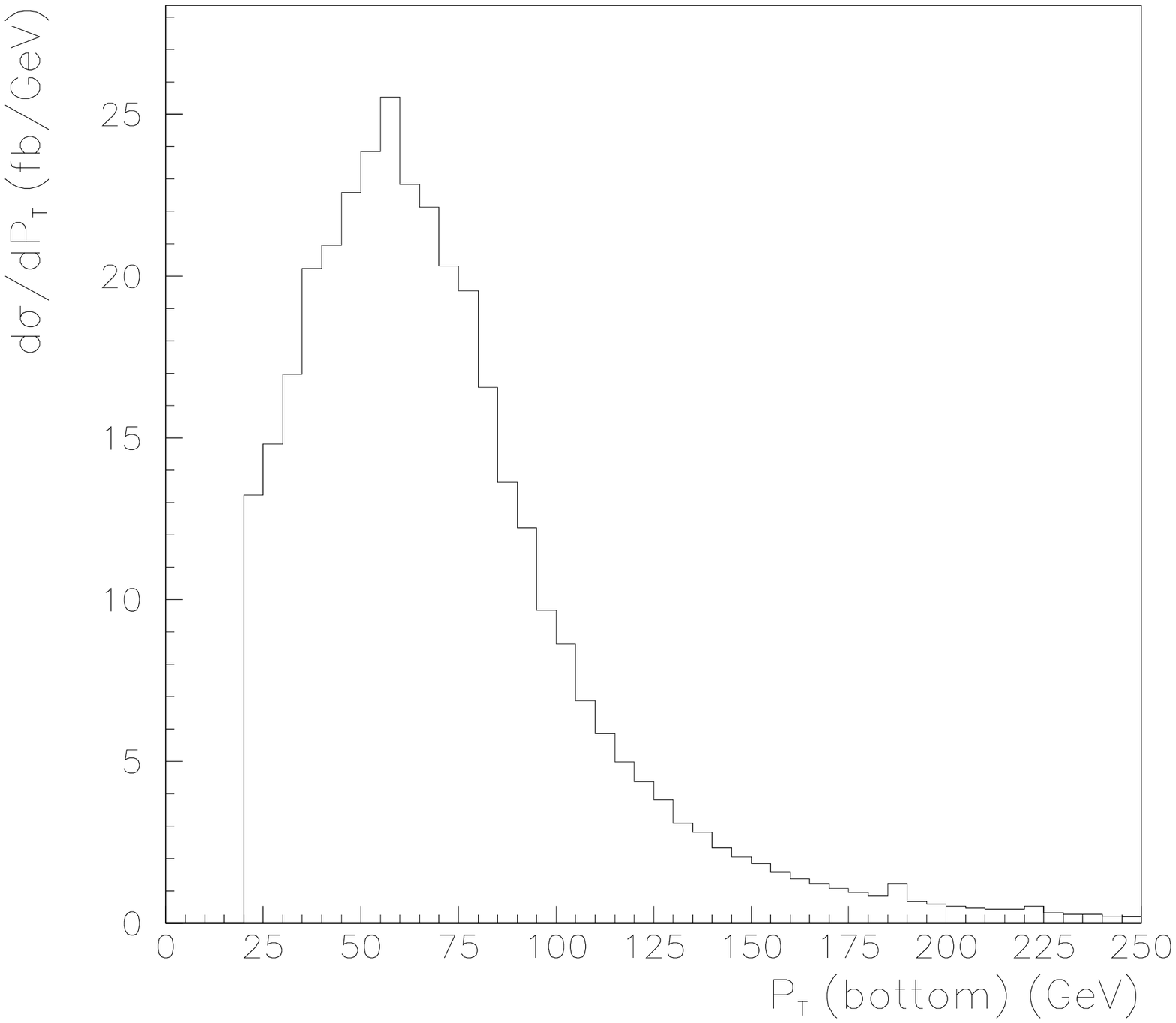}
\end{center}
\par
\caption{Bottom transversal momentum distribution corresponding to single
top production at the LHC. The calculation has been performed at tree level
in the SM ($g_{L}=1$, $g_{R}=0$). }
\label{pbtrans_gl=1_gr=0}
\end{figure}

\begin{figure}[!hbp]
\begin{center}
\includegraphics[width=80mm]{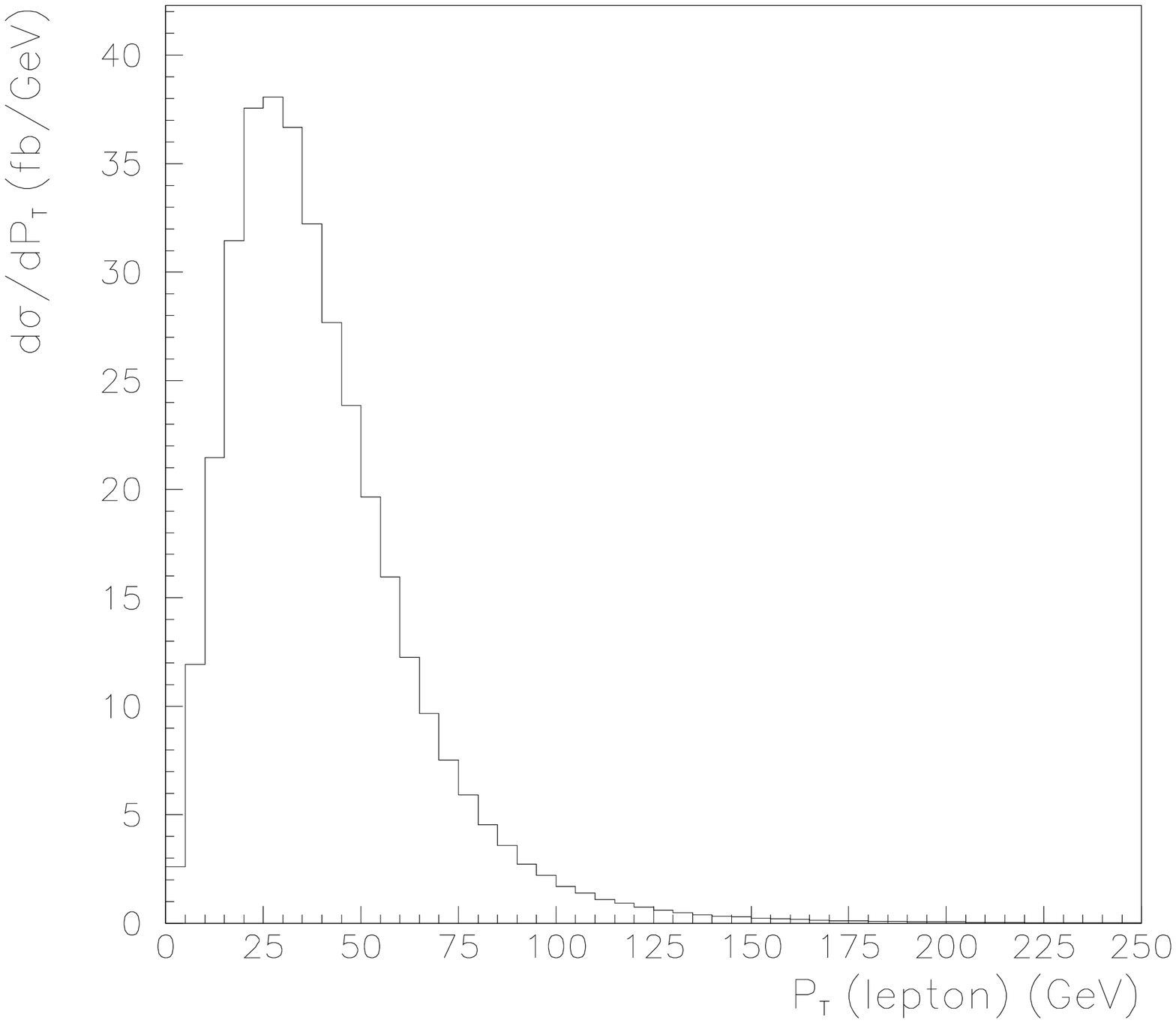}
\end{center}
\par
\caption{Lepton (electron or muon) transversal momentum distribution
corresponding to single top production at the LHC. The calculation has been
performed at tree level in the SM ($g_{L}=1$, $g_{R}=0$). }
\label{pltrans_gl=1_gr=0}
\end{figure}

Fig. \ref{plpbmvv_gl=1_gr=0} shows the invariant
mass of the lepton and bottom system
in the tree level Standard Model. Since we are working in the
narrow width approximation, the distribution falls to zero just below the
physical mass of the top and this reflects the part of the
total momentum of the top carried away
by the undetected neutrino. Figs. \ref{pbtrans_gl=1_gr=0}
and  \ref{pltrans_gl=1_gr=0} actually show the $p_T$ distribution
for the bottom and lepton, respectively, that are produced in the top
decay. As previously discussed, 20 GeV cuts on the respective $p_T$
are imposed. Even though some information
is lost by the fact that the neutrino is not seen and therefore there
is some amount of missing momentum, this does not seem to affect
the sensitivity to the effective couplings too much.
One could as well consider channels in which the $W^+$,
produced in the top decay, decays  hadronically.
In hadronic
decays of the top a full reconstruction of the top mass would be feasible.

Let us now move beyond the Standard Model. Since changing the value
of $g_L$ (while keeping $g_R=0$) amounts to a simple rescaling, we shall
concentrate on the more interesting case of varying $g_R$.
As a rough order-of-magnitude estimate for the effective
$g_R$ coupling we take $\vert g_R\vert=5 \times 10^{-2}$. This is still
worse than the limit implied by $b\to s\gamma$, but is the
sort of sensitivity that LHC will be able to set.
The effects are linear in $g_R$, so it is easy to
scale up or down the results. We have consider the possibility
of $g_R$ having a phase and, accordingly, the experimental sensitivity
to that phase.

\begin{figure}[!hbp]
\begin{center}
\includegraphics[width=13cm]{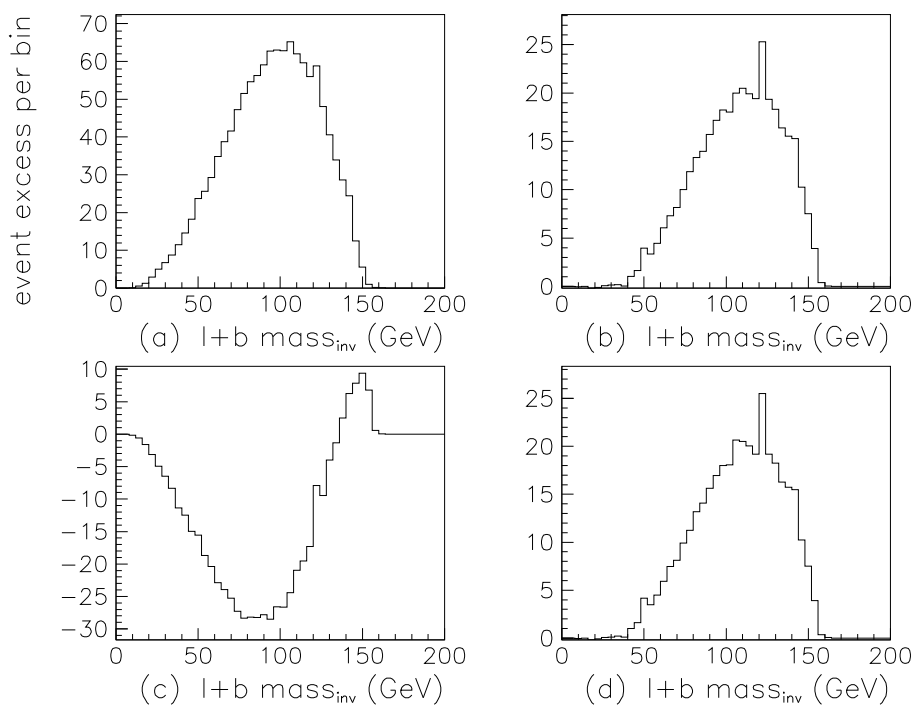}
\end{center}
\par
\caption{Event production difference between non-vanishing $g_{R}$ coupling
caculations and the tree level SM ones ($g_{R}=0$). Differences are plotted
versus the invariant mass of the lepton (electron or muon) plus bottom
system arising in top decay from single top production at the LHC. We have
taken $g_{R}=+5\times 10^{-2}$, $+i5\times 10^{-2},$ $-5\times 10^{-2}$ and $%
-i5\times 10^{-2}$ in plots (a), (b), (c) and (d) respectively.
With the present set of cuts, the total number of events in the
Standard Model is 181,000. The total excess is 1,200, roughly 1\% .}
\label{plpbmvv_gl=1_gr=+-5.d-2_diff}
\end{figure}

\begin{figure}[!hbp]
\begin{center}
\includegraphics[width=130mm]{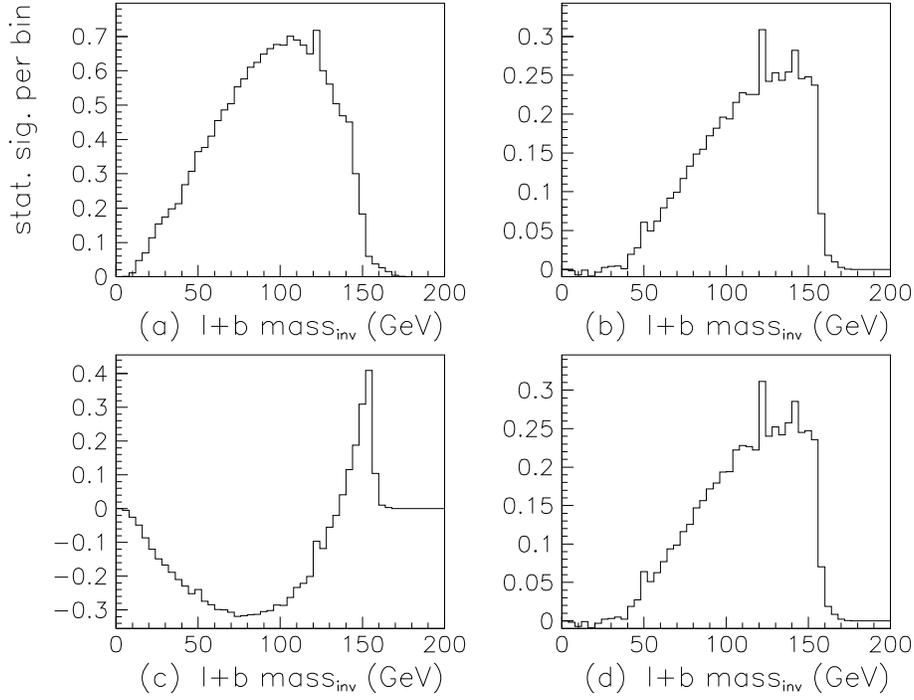}
\end{center}
\par
\caption{Plots corresponding to differences (a), (b) (c) and (d) of Fig. (%
\ref{plpbmvv_gl=1_gr=+-5.d-2_diff}) divided by the square root of the event
production per bin at LHC. The square of the quotient denominator can be
obtained from Fig. (\ref{plpbmvv_gl=1_gr=0}) multiplying $d\protect\sigma
/dm_{inv}$ by the LHC 1-year full luminosity (100 $\mathrm{fb}^{-1}$) and by
the width of each bin (4 GeV. in Fig. (\ref{plpbmvv_gl=1_gr=0})). Taking the
modulus of the above plots we obtain the statistical significance of the
corresponding signals per bin. Note that statistical significance has a
strong and non-linear dependence both on the invariant mass and the right
coupling $g_{R.}$ However purely imaginary couplings are almost insensible
to their sign.}
\label{plpbmvv_gl=1_gr=+-5.d-2_ss}
\end{figure}

We have found that the anti-lepton plus bottom invariant mass
distribution we just discussed in the previous paragraph is
actually sensitive to $g_{R}$. Figs.
\ref{plpbmvv_gl=1_gr=+-5.d-2_diff} and
\ref{plpbmvv_gl=1_gr=+-5.d-2_ss} reflect this sensitivity with the
second figure showing the statistical significance per bin.

\begin{figure}[!hbp]
\begin{center}
\includegraphics[width=130mm]{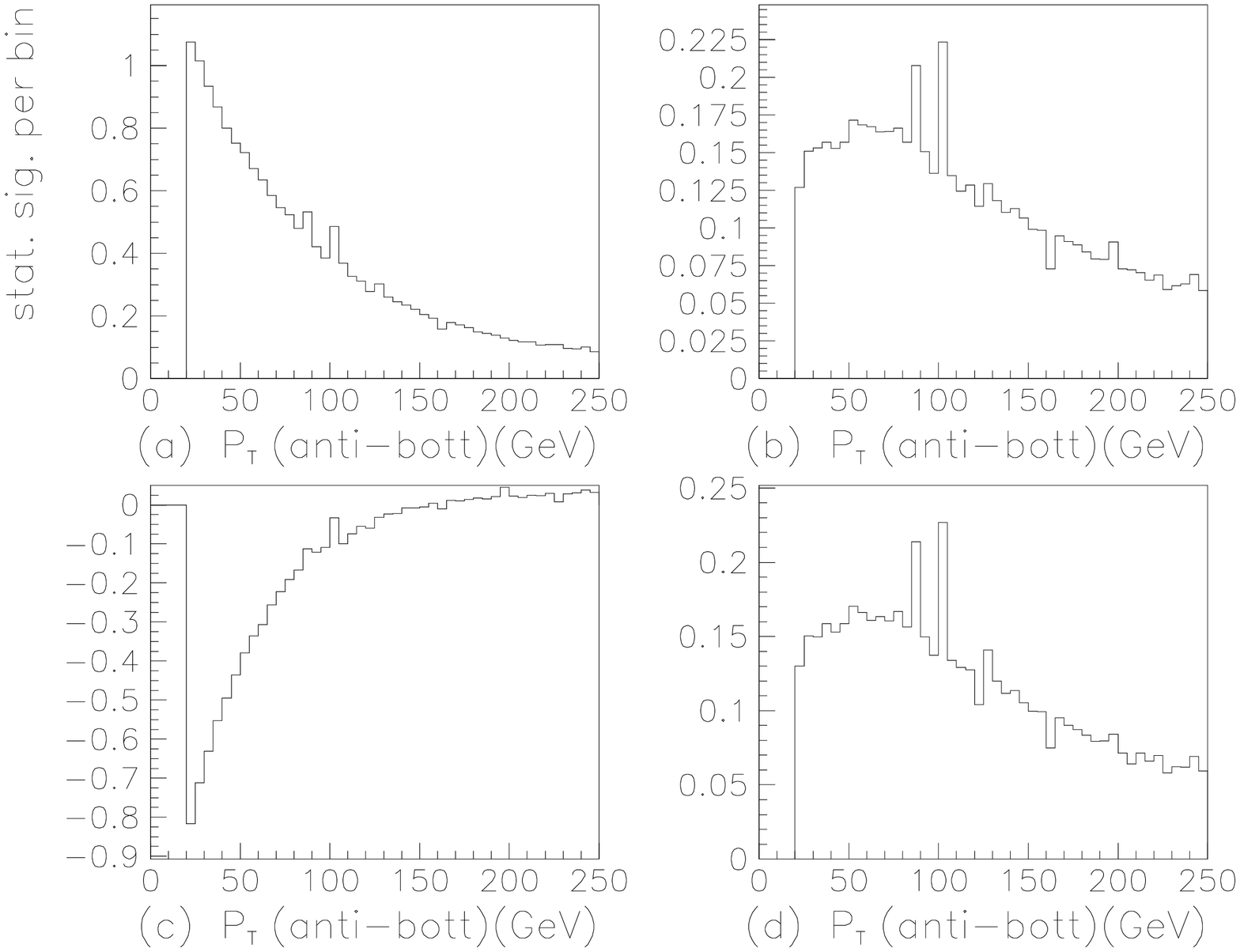}
\end{center}
\par
\caption{Statistical
significance per bin with respect to
anti-bottom transversal momentum. Like in Fig. (\ref
{plpbmvv_gl=1_gr=+-5.d-2_ss}) we have taken $g_{R}=+5\times 10^{-2}$, $%
+i5\times 10^{-2},$ $-5\times 10^{-2}$ and $-i5\times 10^{-2}$ in plots (a),
(b), (c) and (d) respectively. Note that here statistical significance has a
strong dependence on the anti-bottom transversal momentum but is almost
linear on $\emph{Re}\left( g_{R}\right) $ and almost insensible to the sign
of $\emph{Im}\left( g_{R}\right) $. }
\label{pabtrans_gl=1_gr=+-5.d-2_ss}
\end{figure}

\begin{figure}[!hbp]
\begin{center}
\includegraphics[width=130mm]{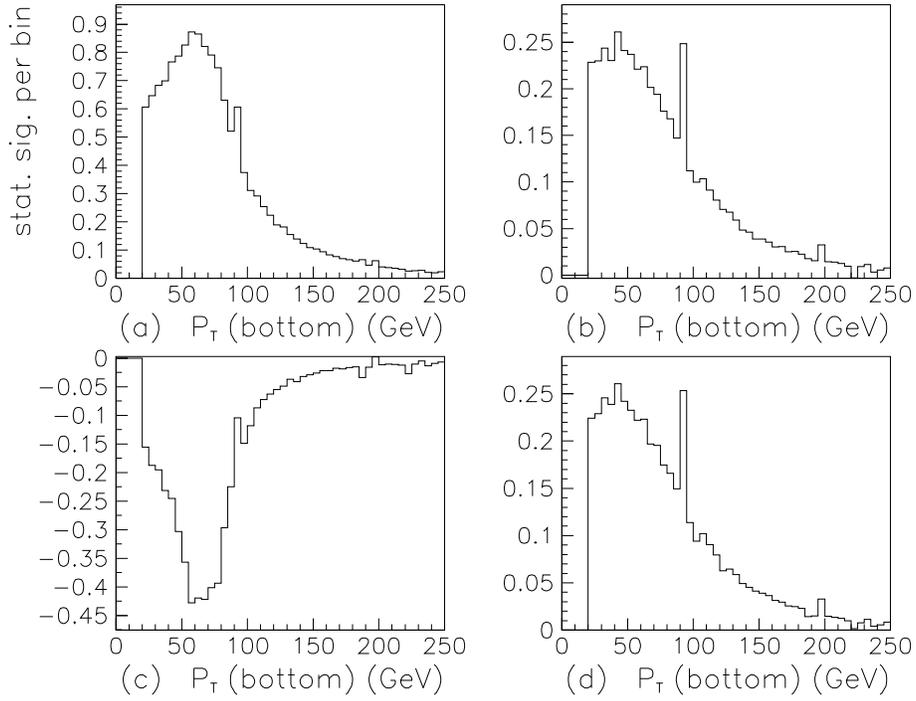}
\end{center}
\par
\caption{Like in Fig. (\ref{plpbmvv_gl=1_gr=+-5.d-2_ss}) we
have taken $g_{R}=+5\times 10^{-2}$, $+i5\times 10^{-2},$ $-5\times 10^{-2}$
and $-i5\times 10^{-2}$ in plots (a), (b), (c) and (d) respectively. Note
that here statistical significance has a strong dependence on the bottom
transversal momentum and clearly favors positive values of $\emph{Re}\left(
g_{R}\right) $ and again is insensible to the sign of $\emph{Im}\left(
g_{R}\right) $. }
\label{pbtrans_gl=1_gr=+-5.d-2_ss}
\end{figure}

\begin{figure}[!hbp]
\begin{center}
\includegraphics[width=130mm]{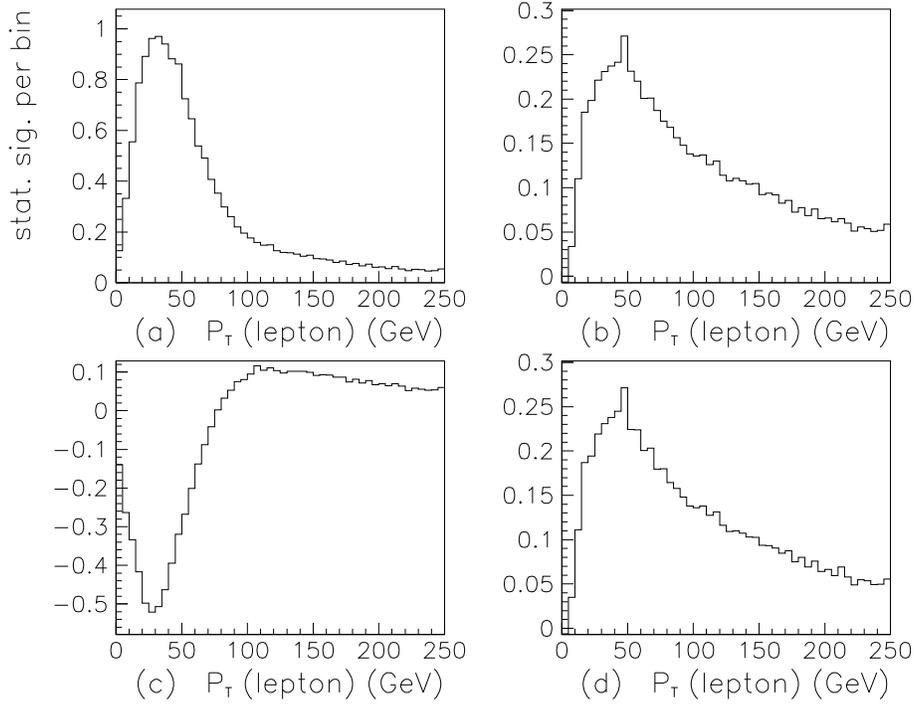}
\end{center}
\par
\caption{Statistical
significance of the corresponding signals per bin with respect to lepton
(electron or muon) transversal momentum. Like in Fig. (\ref{plpbmvv_gl=1_gr=+-5.d-2_ss})
we have taken $g_{R}=+5\times 10^{-2}$, $%
+i5\times 10^{-2},$ $-5\times 10^{-2}$ and $-i5\times 10^{-2}$ in plots (a),
(b), (c) and (d) respectively. Note that again statistical significance has
a strong dependence on the lepton transversal momentum and clearly favors
positive values of $\emph{Re}\left( g_{R}\right) .$ The sign of $\emph{Im}%
\left( g_{R}\right) $ cannot be distinguished.}
\label{pltrans_gl=1_gr=+-5.d-2_ss}
\end{figure}

We shall now show the dependence of the three $p_T$ distributions ($b$,
$\bar{b}$ and lepton) to the modulus and phase of the effective coupling
$g_R$. In all cases the value $g_L=1$ is taken.
The sensitivity to departures from the tree level SM is shown in
Figs. (\ref{pabtrans_gl=1_gr=+-5.d-2_ss}), (\ref{pbtrans_gl=1_gr=+-5.d-2_ss}%
) and (\ref{pltrans_gl=1_gr=+-5.d-2_ss}).

\begin{figure}[!hbp]
\begin{center}
\includegraphics[width=150mm]{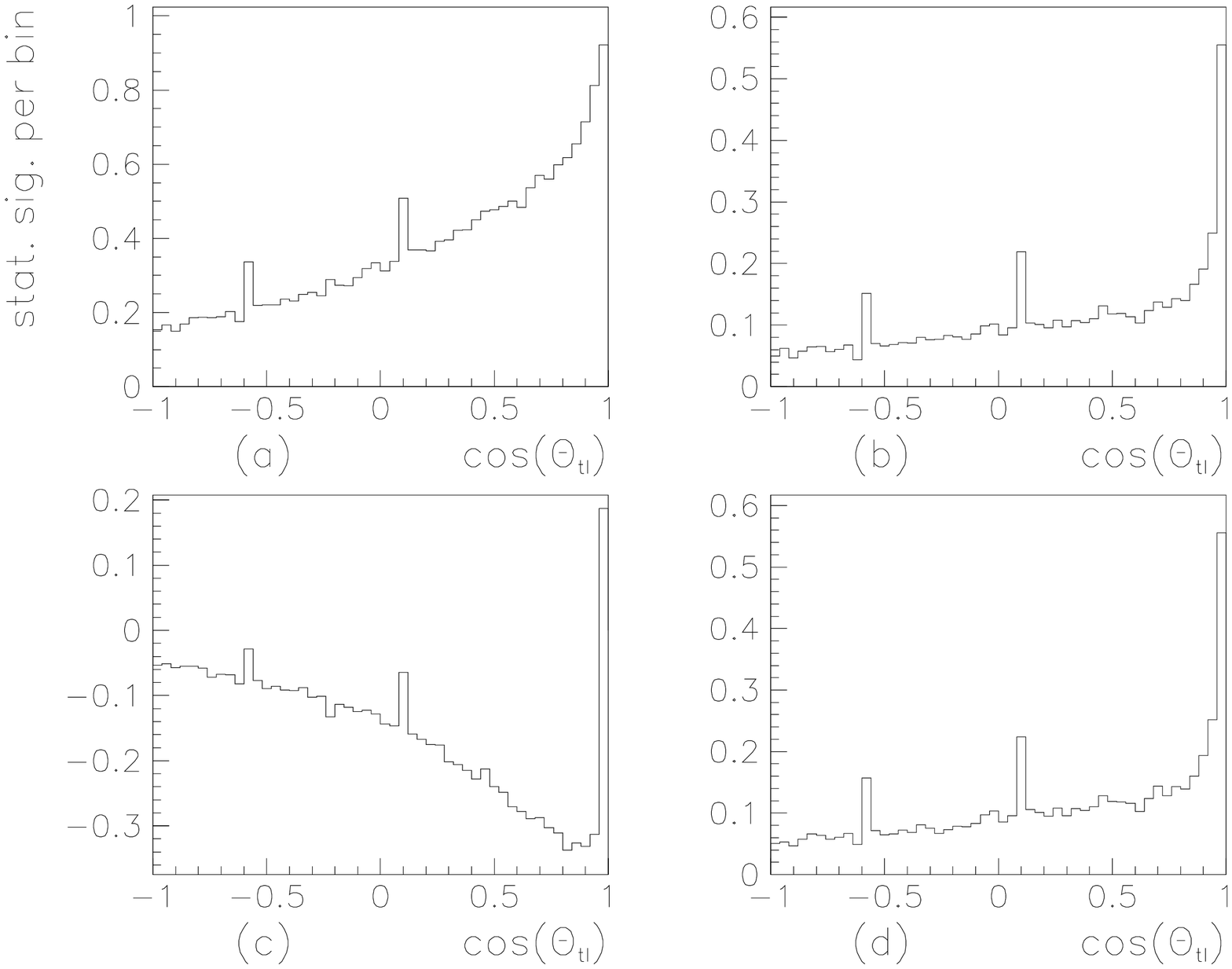}
\end{center}
\par
\caption{Statistical
significance of the corresponding number of events
 per bin with respect to $\cos
\left( \protect\theta _{tl}\right) =$ $\vec{p}_{l}\cdot \left( \vec{p}_{l}+%
\vec{p}_{b}\right) $ $/$ $\left| \vec{p}_{l}\right| \left| \vec{p}_{l}+\vec{p%
}_{b}\right| $ where $\vec{p}_{l}$ and $\vec{p}_{b}$ are
respectively the tree momenta of the lepton (positron or
anti-muon) and bottom. The combination $\vec{p}_{l}+\vec{p}_{b}$
is the best experimental reconstruction of the top momentum
provided the neutrino information is
lost. Like in Fig. (\ref{plpbmvv_gl=1_gr=+-5.d-2_ss}) we have taken $%
g_{R}=+5\times 10^{-2}$, $+i5\times 10^{-2},$ $-5\times 10^{-2}$ and $%
-i5\times 10^{-2}$ in plots (a), (b), (c) and (d) respectively. Note that
again statistical significance has a strong dependence on $\cos \left(
\protect\theta _{tl}\right) $.}
\label{cosptnpl_gl=1_gr=+-5.d-2_ss}
\end{figure}

\begin{figure}[!hbp]
\begin{center}
\includegraphics[width=150mm]{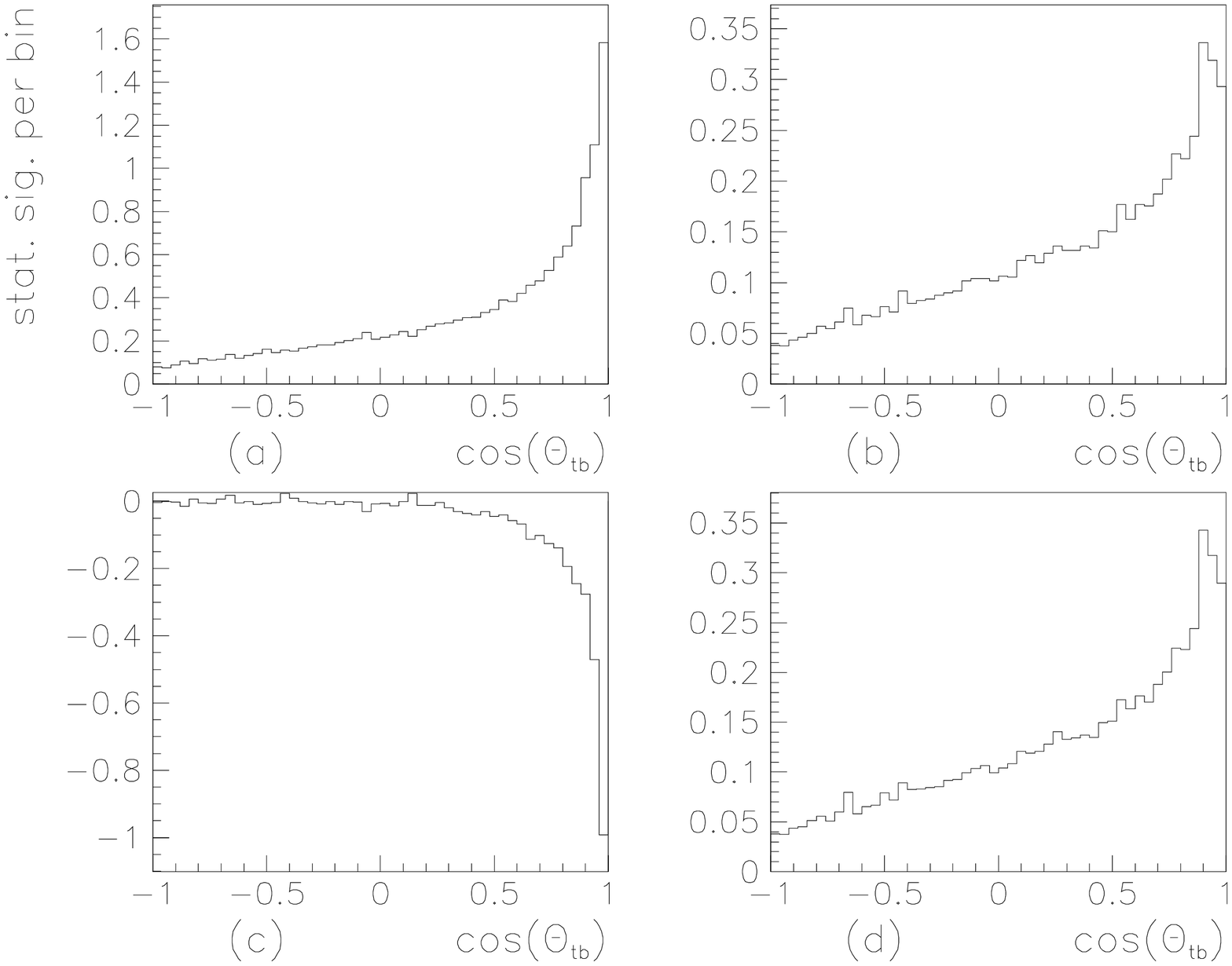}
\end{center}
\par
\caption{Statistical
significance per bin with respect to $\cos
\left( \protect\theta _{tb}\right) =$ $\vec{p}_{b}\cdot \left( \vec{p}_{l}+%
\vec{p}_{b}\right) $ $/$ $\left| \vec{p}_{l}\right| \left| \vec{p}_{l}+\vec{p%
}_{b}\right| $ where $\vec{p}_{l}$ and $\vec{p}_{b}$ are
respectively the tree momenta of the lepton (positron or
anti-muon) and bottom. The combination $\vec{p}_{l}+\vec{p}_{b}$
is the best experimental reconstruction of the top momentum
provided the neutrino information is
lost. Like in Fig. (\ref{plpbmvv_gl=1_gr=+-5.d-2_ss}) we have taken $%
g_{R}=+5\times 10^{-2}$, $+i5\times 10^{-2},$ $-5\times 10^{-2}$ and $%
-i5\times 10^{-2}$ in plots (a), (b), (c) and (d) respectively. Note that
again statistical significance has a strong dependence on $\cos \left(
\protect\theta _{tb}\right) $.}
\label{cosptnpb_gl=1_gr=+-5.d-2_ss}
\end{figure}

We also include the statistical
significance per bin for the signal vs $\cos \left( \theta _{tl}\right) $ in
Fig. (\ref{cosptnpl_gl=1_gr=+-5.d-2_ss}) and vs $\cos \left( \theta
_{tb}\right) $ in Fig. (\ref{cosptnpb_gl=1_gr=+-5.d-2_ss}). $\cos \left(
\theta _{tl}\right) $ and $\cos \left( \theta _{tb}\right) $ are the cosines
of the angle between the best reconstruction of top momentum and the momenta
of anti-lepton and bottom, respectively. In these figures we can clearly see
that low angles corresponds to bigger sensitivities. This is in qualitative
accordance with Eq. (\ref{xurro}) which, after inspection,
tells us that anti-leptons are
predominantly produced in the direction of the top spin and therefore most
of those produced predominantly in the top direction come from a top mainly
polarized in a positive helicity state. Thus the quantity of those
anti-leptons is more sensitive to variations in $g_{R}.$ Even though this
argument applies in the top rest frame, the fact that most of the kinematics
lies in the beam direction makes it valid at least for this kinematics. With
the cuts considered here, the Standard Model
 prediction at tree level for the total
number of events at LHC with one year full luminosity is $180700$. Using
the values $g_{L}=1,$ $g_{R}=+5\times 10^{-2}$ leads to an excess of $1220$
events which corresponds to a $2.9$ standard deviations signal. The $%
g_{L}=1,$ $g_{R}=-5\times 10^{-2}$ model has a deficit of $480$ events which
corresponds to a $1.1$ standard deviations signal. Finally the $g_{L}=1,$ $%
g_{R}=\pm i5\times 10^{-2}$ model has an excess of $367$ events which
corresponds to a $0.86$ standard deviations. We see that there is a large
dependence on the phase of $g_{R}.$

It is perhaps interesting to remark that after considering the
decay process, the sensitivity to $g_R$ is actually quite comparable
to the one obtained in the $t$-channel, where it was assumed that
the polarized top was observable. From this point of view, not much
information gets diluted through the process of top decay.

The implementation of careful selected cuts can slightly improve
these statistical significances but since here we are interested
in an order of magnitude estimate we will not enter into such
analysis here. Moreover since backgrounds are bound to worsen the
sensitivity the above results must be taken as order of magnitude
estimates only. A more detailed analysis goes beyond the scope of
this article.


\section{Conclusions}

In this paper we have performed a full analysis of the sensitivity of
single top production in the \textit{s}-channel to the presence of effective
couplings in the effective electroweak theory. The analysis has been done in
the context of the LHC experiments. We have implemented a set
of cuts which is appropriate to general-purpose experiments
such as ATLAS or CMS. The study complements the one presented in \cite{EM}
that was devoted to the dominant $t$-channel process.

We have seen that the determination of the right effective coupling
in such an experimental context is quite challenging. One has to include
both polarization effects and $m_b$ corrections. Analytical formulae
are presented.

Unlike in the discussion concerning the single top production through the
dominant \textit{t}-channel, top decay has been considered. The only
approximation involved is to consider the top as a real particle (narrow
width approximation).

We have paid careful attention to the issue of the top polarization. We have
argued, first of all, why it is not unjustified to neglect the interference
term and to proceed as if the top spin was determined at an intermediate
stage. We have provided a spin basis where the interference term is
minimized. A similar analysis applies to the \textit{t}-channel process. We
present here and explicit basis for this case too. We get a sensitivity to $%
g_{R}$ in the same ballpark as the one obtained in the \textit{t}-channel
(where decay was not considered). Finally we have obtained that observables
most sensible to $g_{R}$ are those where anti-lepton and bottom momenta are
cut to be almost collinear.

\section*{Acknowledgements}

It is a pleasure to thank G.D'Ambrosio and F.Teubert for detailed
discussions concerning the manuscript. We also acknowledge fruitful
early conversations with M.J.Herrero and J.Fernandez de Troconiz.
We thank D.Peralta for technical help. 
D.E. wishes to thank the hospitality of the CERN TH Division, where this
work was finished.
J.M acknowledges the support from
Generalitat de Catalunya, grant 1998FI-00614. Financial support from grants
FPA2001-3598, 2001SGR 00065 and the EURODAPHNE network are also
acknowleged.

\end{document}